\documentclass[a4paper,11pt]{article}
\usepackage[a4paper,left=2.73cm,right=2.7cm,top=3cm,bottom=3.5cm]{geometry}

\usepackage{amsfonts,amsmath,amssymb,tabstackengine,amsthm}
\usepackage{array}
\usepackage{comment}
\usepackage{mathrsfs}
\usepackage[bb=boondox]{mathalfa}
\usepackage{cite}

\usepackage{MnSymbol}
\usepackage{tensor}

\usepackage[colorlinks=true,linktocpage=true,linkcolor=blue,citecolor=blue]{hyperref}
\usepackage{graphicx}

\addtocontents{toc}{\protect\setcounter{tocdepth}{2}}
\numberwithin{equation}{section}

\newtheorem{theorem}{Theorem}

\theoremstyle{definition}
\newtheorem{definition}{Definition}
\newtheorem*{example}{Example}
\newtheorem*{remarks}{Remarks}

\def\cD{{\cal D}}

\def\cN{{\cal N}}

\def\Es{{\mathrm{E}_{7(7)}}}
\def\es{{\mathfrak{e}_{7(7)}}}
\def\Ed{{\mathrm{E}_{d(d)}}}
\def\EdR{{\mathrm{E}_{d(d)}\times \mathbb{R}^+}}
\def\ed{{\mathfrak{e}_{d(d)}}}
\def\SL{{\mathrm{SL}}}

\def\SU{{\mathrm{SU}}}
\def\SO{{\mathrm{SO}}}
\def\GL{{\mathrm{GL}}}
\def\Mint{{M_{\text{int}}}}
\def\Hprin{H_{\text{prin}}}
\def\Mloc{{M_{\text{loc}}}}

\newcommand{\scalP}[2]{{\langle #1 ,\,#2 \rangle}}

 \usepackage [latin1]{inputenc}

\begin{document}
\begin{titlepage}

\thispagestyle{empty}

\begin{center}
{\LARGE \textbf{How to uplift non-maximal gauged supergravities}}

\vspace{40pt}
		
{\large \bf Davide Rovere}$\,^{a,b}$  \,\,\,  \large{and} \,\,\, {\large \bf Colin Sterckx}$\,^{c}$
		
\vspace{25pt}

$^a$\,{\normalsize 
Dipartimento di Fisica e Astronomia ``Galileo Galilei'', \\
Universit\`a di Padova, Via F. Marzolo 8, 35131, Padua, Italy
}
\\[10mm]

$^b$\,{\normalsize 
INFN, Sezione di Padova, Padua, Italy
}
\\[10mm]

$^c$\,{\normalsize
Universit\'e Libre de Bruxelles (ULB) and International Solvay Institutes,\\
Service  de Physique Th\'eorique et Math\'ematique, \\
Campus de la Plaine, CP 231, B-1050, Brussels, Belgium.}
\\[7mm]

\texttt{davide.rovere@studenti.unipd.it}  \,\, , \,\, \texttt{colin.sterckx@ulb.be}

\vspace{40pt}

\abstract{
\noindent In this paper, we provide an algorithm to perform the uplift of non-maximal $G_g$-gauged supergravities to type IIB or 11D supergravities. Using tools of exceptional field theory and generalised geometry, we show that the internal manifold admits a $G_g$-action, and that consistency of the uplift is equivalent to solving a simpler PDE on the quotient $M_{\text{int}}/G_g$. As an application, we classify all possible uplifts of pure half-maximal four-dimensional $\textrm{SO}(4)$-gauged supergravity to type IIB and we recover consistent truncations around any of the D'Hoker-Estes-Gutperle solutions \cite{DHoker:2007hhe}.}

\end{center}
\end{titlepage}

\tableofcontents

\hrulefill
\vspace{10pt}

\section{Introduction}
\label{sec:Intro}

Over the last decades, various approaches have been developed to study and classify solutions to the equations of motion of ten- and eleven-dimensional supergravities. Amongst these, \emph{consistent truncations} play a central role. This method fixes some of the dynamical fields of the original supergravity to a specific ansatz such that the equations of motion of the full theory reduce to those of a simpler lower-dimensional \emph{gauged} supergravity (gSUGRA). These truncations are called ``\emph{consistent}" if any solution of the reduced theory uplifts to a solution of the full theory, thereby providing a higher-dimensional embedding, relevant for their string or M-theory interpretation.

Constructing consistent truncations is a longstanding and notoriously difficult problem due to the non-linearity of the supergravity equations of motion. In the vast majority of examples, the consistency of the truncation is established by means of a symmetry argument, and even apparent exceptions were later reinterpreted in this framework\footnote{E.g. the consistent truncation of \cite{Bobev:2023bxs} is an example of the type of consistent truncations later built in \cite{Guarino:2024gke}.}. The basic idea, due to \cite{Coleman1985App}, is to retain all modes invariant under a symmetry group of the higher-dimensional theory, and nothing else. This ensures that no truncated mode can be sourced by an invariant one, which implies consistency. Early examples in the context of supergravity, generalising the construction of Kaluza and Klein, were tori compactifications yielding reductions to \emph{ungauged} lower-dimensional supergravities \cite{Cremmer:1997ct}. In that setup, the consistency of the truncation relied on a $\mathrm{U}(1)^d$ symmetry group, corresponding to translation along the internal manifold $T^d$. Scherk and Schwarz \cite{Scherk:1978ta,Scherk:1979zr} later extended this reasoning when the internal manifold is a Lie group by keeping modes invariant under the left (or right) action of the Lie group on itself. The consistency of the truncation can be reframed as the existence of a parallelisation on the internal manifold with constant torsion, identified with the underlying Lie algebra structure constant.

    Despite these progresses, proving the consistency for truncations on generic \emph{coset spaces} -- relevant e.g. for the Freund-Rubin solutions AdS$_5 \times S^5$ in type IIB, or AdS$_4\times S^7$ and AdS$_7\times S^4$ in M-theory -- remained elusive for decades. The advent of \emph{Exceptional Generalised Geometry} (EGG) \cite{Coimbra:2011nw,Coimbra:2012af} and \emph{Exceptional Field Theory} (ExFT) \cite{Hohm:2013vpa,Hohm:2013uia,Hohm:2014fxa} provided the tools to solve this problem (see \cite{Aldazabal:2013sca,Baguet:2015xha,Berman:2020tqn,Sterckx:2024vju,Samtleben:2025fta} for reviews). They reorganise the degrees of freedom of maximal supergravities to reveal a hidden $\Ed$ symmetry. Within this framework many coset spaces, including the spheres $S^4,\,S^5$, and $S^7$, were shown to admit a generalised notion of parallelisation \cite{Lee:2014mla}. In particular, this can be used to prove the consistency of the truncations of type IIB on $S^5$ to the $\SO(6)$-gauged maximal $D=5$ supergravity, and M-theory to the $\SO(8)$ $D=4$ and $\SO(5)$ $D=7$ gauged supergravities\footnote{Partial uplift of the $D=4$ $\mathcal{N}=8$ $\SO(8)$-gauged supergravity to M-theory had already been worked out in \cite{deWit:1986oxb}, and completed in \cite{Varela:2015ywx}, without the use of ExFT.}. By construction, these ``generalised Scherk-Schwarz" (gSS) \cite{Hohm:2014qga} reductions always reduce to \emph{maximal} gauged supergravities. Examples of such truncations include compactifications on spheres and hyperboloid \cite{Baguet:2015sma} ($D=5$), product thereof \cite{Inverso:2016eet} ($D=4$), and more generic manifolds \cite{Malek:2015hma,Malek:2017cle} ($D=4$ and $D=7$). Upon making suitable adjustments to the ExFT framework, one can also reinterpret consistent truncations from massive type IIA (mIIA) on $S^6$ \cite{Guarino:2015jca,Guarino:2015vca} in a gSS manner \cite{Ciceri:2016dmd,Cassani:2016ncu}. Finally, $D=3$ examples include \cite{Eloy:2021fhc,Eloy:2023zzh} which are induced by the AdS$_3\times S^3 \times S^3\times S^1$ and AdS$_3\times S^3\times T^4$ solutions of type IIB supergravity. 

Further refinement was achieved in\cite{Cassani:2019vcl} where the authors consider an internal manifold admitting a \emph{generalised} $G_S$-structure, rather than a full generalised parallelisation ($G_S = \{e\}$). They show that the existence of a $G_S$-structure with constant and $G_S$-invariant intrinsic torsion implies the existence of a consistent truncation to \emph{non-maximal} gauged supergravities. In the same paper, these authors worked out several examples of consistent truncations to \emph{half-maximal} gauged supergravities in five dimensions. Amongst other, the type IIB consistent truncation on Sasaki-Einstein manifolds, originating from standard $\SU(2)$-structures \cite{Cassani:2010uw,Gauntlett:2010vu,Liu:2010sa,Skenderis:2010vz}, as well as consistent truncations around the Maldacena-Nunez geometries \cite{Maldacena:2000mw} down to a half-maximal $D=5$ supergravity coupled to three vector multiplets with gauge group $\textrm{U}(1)\times \textrm{ISO}(3)$\footnote{This consistent truncation was also obtained by other means in \cite{MatthewCheung:2019ehr}.}. This last result was further extended in \cite{Cassani:2020cod} and a classification of upliftable $D=5$ $\mathcal{N}=2$ supergravities was performed in \cite{Josse:2021put}. Consistent truncations to non-maximal four-dimensional supergravities are sparse, including the uplift of minimal \cite{Larios:2019lxq} and non-minimal \cite{Cassani:2012pj} $\mathcal{N}=2$ gSUGRA in M-theory, consistent truncation of M-theory on coset spaces \cite{Duboeuf:2022mam,Duboeuf:2023dmq}, $G_2$ invariant subsectors of M-theory \cite{Duboeuf:2024tbd}, and an uplift of minimal $\mathcal{N}=4$ gSUGRA to type IIB \cite{Guarino:2024gke}. We also mention the recent line of research concerning exceptional generalized cosets \cite{Hassler:2023axp,Hassler:2025qhh} also aiming at building uplifts of non-maximal gauged supergravity.

Amongst the many questions regarding a possible classification of consistent truncations, we highlight two relevant for our work.
\begin{itemize}
    \item \textbf{The ``top-down" problem:} Given a solution of eleven-dimensional or type II supergravity on $M_{\text{ext}}\times M_{\text{int}}$, is there a consistent truncation which includes this solution \emph{and} reduces the full equations of motion to those of a gauged supergravity on $M_{\text{ext}}$?\\[1mm]
    The answer is yes for supersymmetric AdS$_D$ and Mink$_D$ solutions. As conjectured in \cite{Gauntlett:2007ma} and proven in \cite{Cassani:2019vcl}, $\mathcal{N}$-supersymmetric AdS$_D$ solutions define a consistent truncation to the corresponding \emph{pure} gauged supergravity. This implies a notion of universal sub-sectors common to any $\mathcal{N}$-supersymmetric AdS$_D$ solutions. This makes sense since this sub-sector is dual to the stress-energy supermultiplet of the dual CFT$_{D-1}$. However, nothing guarantees that such a truncation exists in the absence of supersymmetry or if one requires the presence of additional matter multiplet in the reduced theory.
    
    \item \textbf{The ``bottom-up" problem:} Given a gauged supergravity in $D$ dimensions, can we embed it consistently in 11D/type II supergravity? \\[1mm]
    If the gauged supergravity is \emph{maximal} ($D\geq 3$), this is precisely the problem treated in \cite{Inverso:2017lrz,Inverso:2024xok}.  The ungauged maximal supergravities admit an $\Ed$ symmetry group, a subgroup $G_g\subset\Ed$ can be gauged by specifying an embedding tensor. When an uplift exists, these gauged supergravities correspond to consistent truncations on a homogeneous manifold of the form  $G_g/H \times U(1)^r$ and a reduction ansatz can be explicitly built.  
\end{itemize}
As such, most examples and explicit constructions remain confined to uplifts of \emph{maximal} gauged supergravities and truncations to \emph{pure} supergravities. In this work, we refine the general techniques used to uplift maximal gauged supergravities using ExFT/EGG, extending the gSS construction beyond maximal/pure SUGRA setting. 

\bigskip

\paragraph{How to uplift a gauged supergravity?} Starting from a $D$-dimensional gauged supergravity, we explain how to characterise its possible uplift(s). A gauged supergravity is characterised by its matter content and by an embedding tensor $\Theta$, specifying its gauge group $G_g$. We show that it can only be uplifted to 11D/type IIB supergravity if the internal manifold admits a $\mathfrak{g}_g$-action. This allows us to foliate $\Mint$ in $G_g$-orbits. Although the quotient $\Mint/G_g$ might be singular, the internal manifold, locally and away from singularities, can be modelled as
\begin{equation}
    \Mint \approx G_g/H \times B \,.
    \label{eq:MintIntro}
\end{equation}
The ``base" $B$ is a topologically trivial manifold of the appropriate dimension, while $H$ is a subgroup of $G_g$. This subgroup controls whether an uplift, when it exists, is in type IIB or M-theory. Classifying consistent truncations is then equivalent to the classification of generalised $G_S$-structures on manifolds of the type $\Mint$ in \eqref{eq:MintIntro} with an intrinsic torsion equal to the embedding tensor $\Theta$. 

There are two ways to specify a generalised $G_S$-structure, either in terms of $G_S$-invariant sections of the generalised bundle, or in term of a globally well defined frame. We use both to obtain different information about the $G_S$-structure. When it is defined in terms of $G_S$-invariant sections $K_A$ of the generalised tangent bundle on $\Mint$, we show that the $K_A$ must be built out of (co-)closed $G_g$-equivariant poly-forms. This allows us to provide a simple ansatz for the most generic frame $E$ defining the $G_S$-structure. This section of $G_S\backslash \Ed$ is built by solving a series of algebraic equations related to the section constraint of ExFT. Finally, we show that enforcing the torsion condition on our frame ansatz reduces to a simpler set of PDE not on the full $\Mint$ but only on $B$.

We do make mild topological assumptions on $G_g$ and $G_g/H$. We assume that the $\mathfrak{g}_g$-action lifts to a \emph{proper} $G_g$-action on $\Mint$. Depending on the type of uplift, we assume that the de Rham cohomology groups $H^p_{dR}(G_g/H) =0$ for $p=1,\,3,\,5,\,7$ in type IIB or $p=4,\,7$ in M-theory. Finally, we do not consider uplifts to massive type IIA supergravity, which we leave for future work.

\paragraph{Plan of the paper}
In section \ref{sec:Review}, we review the aspects of ExFT, generalised $G$-structures, and intrinsic torsion relevant to our work. In section \ref{sec:Construction}, we show how to build the uplift of any given supergravity, when it exists. In section \ref{sec:nv6Example}, we uplift an $\mathcal{N}=4$ $D=4$ $n_v=6$ gauged supergravity in M-theory. This example serves to illustrate our method (although the resulting uplift is relatively trivial). Finally, in section \ref{sec:TypeIIBExample}, we classify the possible uplifts of pure $\mathcal{N}=4$ $D=4$ gauged supergravity in type IIB. We compute explicitly the supergravity fields, by applying the SUGRA/ExFT dictionary and we obtain the consistent truncations around any of the solutions of \cite{DHoker:2007hhe}. We included several appendices relevant for our computations. In appendix \ref{app:Conventions}, we collect our conventions for naming indices. Appendix \ref{app:GroupActions} summarises relevant mathematical results concerning Lie group actions on manifold as can be read in \cite{Meinrenken2003}. In appendix \ref{app:Equiv}, we condense relevant information on the classification of equivariant forms. The appendix \ref{app:EflatConstraints} presents several proofs constraining our ansatz for the frame which were omitted from the main text for legibility. Finally, \ref{app:KKCalculus} compiles the tools needed to check the equations of motions and the gauge invariance of the uplift. It also includes the non-linear field redefinitions needed to build the $\Es$-ExFT to type IIB dictionary.

\section{Review of ExFT and consistent truncations}
\label{sec:Review}

We start here by summarising known results concerning ExFT, consistent truncations, as well as fixing our conventions and notations for the rest of the paper. Since, in this paper, we will work out the uplift of four-dimensional supergravities, we will provide the details for the $\Es$-ExFT. Finally, we give an explicit formulation of the $\Es$-generalised Lie derivative in the language of generalised geometry for type IIB supergravity, previously absent in the literature.

\subsection{Generalised geometry and exceptional field theory}
\label{subsec:Generalities}
To define an ExFT, we consider type IIB or M-theory on a product manifold $M_{\text{ext}} \times \Mint$, where the external manifold $M_{\text{ext}}$ has dimension $D$, and the internal manifold $M_{\text{int}}$ has dimension $d'$. The fundamental idea of ExFT is to encode both the parameters of infinitesimal diffeomorphisms and those of the gauge transformations of the supergravity in the same object: a generalised vector $V^M$. This object is a section of a generalised tangent bundle on $\Mint$, which is schematically a $R_1$-vector bundle on $\Mint$. The vector space $R_1$ denotes the fundamental representation of $\Ed$. It turns out that it is convenient for computations to extend the coordinates $y^m$ on $\Mint$ with auxiliary coordinates $\{Y^M \}\supset \{y^m\}$ and to consider $V^M$ as a vector on $\mathbb{R}^{|R_1|}$. This construction should be thought of as a generalisation of the usual vector fields parametrising infinitesimal diffeomorphisms and transforming under $\GL(d')$. We define the generalised Lie derivative as
\begin{equation}
    L: E \otimes_{\mathbb{R}} E \mapsto E: (\Lambda,\,V) \mapsto (L_\Lambda V)^M := \Lambda^N \partial_N V^M - \alpha_d\, \mathbb{P}^M{}_N{}^P{}_Q \partial_P \Lambda^Q V^N\,,
\end{equation}
where $\alpha_d$ is a constant real number which depends on the rank of $\Ed$, and 
\begin{equation}
\mathbb{P}^M{}_N{}^P{}_Q   = (t_\alpha)^M{}_N (t^\alpha)^P{}_Q
\end{equation}
is the projector on the adjoint representation ($t_\alpha$ is a basis of $\ed$). The generalised infinitesimal diffeomorphisms are defined as $\delta_V = L_V$. The ExFT action must be invariant under these diffeomorphisms, i.e. $\delta_V S_{\text{ExFT}} = 0$.

A characteristic of ExFT is that the generalised infinitesimal diffeomorphisms only close if all the fields satisfy the ``\emph{section constraint}":
\begin{equation}
    Y^{MN}{}_{PQ} \partial_M \otimes \partial_N = 0\,,
\end{equation}
for a tensor $Y^{MN}{}_{PQ}$ whose specifics depend on $d$\cite{Inverso:2017lrz}.  These constraints effectively reduces the auxiliary coordinates $Y^M$ to the physical ones $y^m$ satisfying $Y^{mn}{}_{PQ} =0$. Due to this, we will often understand solutions to the section constraint as given by a tensor $\mathcal{E}_M{}^m$ satisfying
\begin{equation}
    Y^{MN}{}_{PQ}\,\mathcal{E}_M{}^{m}\mathcal{E}_N{}^{n} = 0\,.
    \label{eq:secConstraints}
\end{equation}
such that $\partial_M = \mathcal{E}_M{}^m\partial_m$. For any $d$, there are exactly two distinct and maximal solutions to the section constraint, which correspond to type IIB ($d'=d-1$) and 11D ($d'=d$) supergravity respectively. The type IIB coordinates transform in the $(\mathbf{d-1},\,1)$-representation of $\GL(d-1) \times \SL(2)_{\text{IIB}} \subset \Ed \times \mathbb{R}$, while the 11D coordinates transform under the vector representation of $\GL(d) \subset \Ed \times \mathbb{R}$. Branching $R_1$ in $\GL(d')$ representations, generalised vectors can be identified with poly-forms. This poly-form contains a vector on a $d$ or $(d-1)$ dimensional manifold, parametrising usual infinitesimal diffeomorphisms, as well as a series of $p$-forms densities, parameters of the gauge transformations.

In this language, the generalised Lie derivative can always be rewritten as
\begin{equation}
    (L_\Lambda V)^M = (\mathcal{L}_{\Lambda^m} V)^M + \text{terms in}\,\, (d\Lambda,\,V) \,.
\end{equation}
where $\mathcal{L}$ is the usual Lie derivative acting on $p$-form densities.

Finally, we specify the field content of ExFTs. It always contains a generalised metric $\mathcal{M}(x,\,Y)$, equivalent to a frame $E \in K_d\backslash \EdR$, where $K_d$ is the maximal compact subgroup of $\EdR$. In the supergravity language, this generalised metric encodes degrees of freedom corresponding to the internal metric as well as fluxes on the internal space. The ExFT field content also contains vector fields, $\mathcal{A}^M(x,\,Y)$, in the $R_1$ representation and a series of $p$-forms forming the ``tensor hierarchy". We refer to the various reviews and the original constructions for more details concerning this tensor hierarchy as well as the precise ExFT actions.

\subsubsection{The $\Es$ generalised Lie derivative in type IIB}

We spell out the construction of the generalised Lie derivative in the case $d=7$ with an internal manifold $M$ for the type IIB solutions to the section constraint. The generalised tangent vectors take value in $R_1 = \mathbf{56}$, the fundamental representation of $\Es$. We perform the branching under the $\GL(6)\times \SL(2)_{\mathrm{IIB}}$ subgroup of $\Es$. The $\GL(6)$ group corresponds to the structure group of the internal manifold and $\SL(2)$ is the global symmetry of type IIB supergravity. The branching reads
\begin{equation}
\begin{array}{rcccccccccc}
    E &\cong&\,  TM &\oplus& T^*M_{+1}  &\oplus& T^*M \otimes S &\oplus& TM_{+1} \otimes S^* &\oplus& \Lambda^3 T^*M\,\\[2mm]
    &:=& E^{(1,\,0)}&\oplus& E^{(0,\,1)}_{+1} &\oplus& E^{\alpha\,(0,\,1)} &\oplus& E^{\alpha\,(1,\,0)}_{+1} &\oplus& E^{(0,\,3)} \,.
    \end{array}
\end{equation}
where $E^{(p,\,q)}_\lambda$ is the vector bundle of $(p,\,q)$-polyforms densities of weight $\lambda$,
and $S$ is a vector bundle associated with the fundamental representation of $\SL(2)$, we use the index $\alpha$ to label its vectors.

We can explicitly identify a generalised vector $V^M \partial_M$ with poly-tensor densities of the form
\begin{equation}
\begin{array}{rcccccccccc}
V & = & v &+ &  \tilde{v} & + & b^\alpha &+ & \tilde{b}_\alpha & + & \lambda_{(3)}\,,\\[2mm]
&& \in E^{(1,\,0)} & & \in E^{(0,\,1)}_{+1}  && \in E^{\alpha(0,\,1)} && \in E^{\alpha \,(1,\,0)}_{+1} && \in E^{(0,\,3)}
\end{array}
\end{equation} 
With these notations, the generalised Lie derivative reads:
\begin{equation}
\begin{array}{rll}
    L_V V' =& \mathcal{L}_v V' &\big\} \in E
    \\
    &+ \iota_{\tilde{b'}_\alpha} db^\alpha - \partial_m \tilde{b}^{m}_{\alpha}\, b'^{\alpha} - (\smallstar \lambda'_{(3)})\cdot d\lambda_{(3)}  &\big\}\in E^{(0,\,1)}_{+1}\\
    &- \iota_{v'} db^\alpha &\big\} \in E^{\alpha\,(0,\,1)}\\
    &- v' \, \partial_m \tilde{b}^m_\alpha + (\smallstar d\lambda_{(3)})\cdot b'^\beta \epsilon_{\beta\alpha} - (\smallstar \lambda'_{(3)}) \cdot  db^\beta \epsilon_{\beta\alpha}&\big\} \in E^{\alpha \,(1,\,0)}_{+1}\\
    &- \iota_{v'} d\lambda_{(3)} + 2\, \epsilon_{\alpha\beta} \, b'^\alpha \wedge d b^\beta\,&\big\}\in E^{(0,\,3)}\,.
\end{array}
\label{eq:genLieDerIIB}
\end{equation}
The operator 
\begin{equation}
\smallstar: \Lambda^pT^*M \rightarrow \Lambda^{6-p}TM_{+1}: \omega_{m_1\,\cdots\,m_p} \rightarrow \frac{1}{p!} \epsilon^{m_1\cdots m_p\,n_1\cdots n_{6-p}} \omega_{m_1\cdots m_p}
\end{equation}
denotes the contraction with the normalised six-dimensional Levi-Civita symbol while the dot is the standard contraction
\begin{equation}
    (v \cdot w)^{a_1\cdots a_q}{}_{b_1\cdots b_r} = \frac{1}{p!} v^{a_1 \cdots a_q\, n_1\cdots n_{p}} w_{b_1\cdots b_r\,n_1\cdots n_{p}}\,.
\end{equation}

\subsubsection{The $\Es$ generalised Lie derivative in M-theory}

We review the same construction for the M-theory solution to the section constraint. The generalised tangent bundle splits in $\GL(7)$ representations according to
\begin{equation}
\begin{array}{ccccccccc}
    E & \cong & TM &\oplus& T^*M_{+1}& \oplus& \Lambda^2 T^*M &\oplus& \Lambda^2T M_{+1} \\
    &:=&E^{(1,0)}&\oplus& E^{(0,1)}_{+ 1}&\oplus&E^{(0,2)}&\oplus& E^{(2,0)}_{+1}\,.
\end{array}
\end{equation}
We identify generalised vectors $V^M\partial_M$ with poly-tensor densities as
\begin{equation}
\begin{array}{rccccccc}
    V =& v &\oplus &\tilde{v} &\oplus &\lambda_{(2)} &\oplus &\tilde{\lambda}_{(2)}\,,\\[1mm]
    &\in E^{(1,\,0)} & &\in E^{(0,\,1)}_{+1} &&\in E^{(0,\,2)} && \in E^{(2,\,0)}_{+1} 
\end{array}
\end{equation}
Using these notations, the generalised Lie derivative reads
\begin{align}
\begin{array}{rll}
L_V V' =& \mathcal{L}_v V' & \big\} \in E\\
 & - \iota_{v'} d(\lambda_{(2)}) & \big\} \in \Lambda^2 T^*M\\
 & - \smallstar (\lambda_2'\wedge d\lambda_{(2)}) - \tfrac{3}{2} \iota_{v'} d\smallstar\tilde{\lambda}_{(2)} & \big\} \in \Lambda^2 TM_{+1}\\
 & -\tfrac{1}{2} d\lambda_{mnp} \tilde{\lambda'}^{np} + \partial_n \tilde{\lambda}^{pn} \lambda'_{pm}& \big\} \in T^*M_{+1}\\
\end{array}
\end{align}
which is equivalent to the formulation presented in \cite{Coimbra:2011ky}, up to Hodge dualities.

\subsection{Systematics of consistent truncations}

Ungauged supergravities are labelled by their dimension $D$, by the number of supersymmetric transformations $\mathcal{N}$, and by their field content (labelled by integers $n_h,\,n_v,\,...$, representing the number and type of additional supermultiplets). Exploiting Hodge dualities, and using a series of field redefinitions, one can show that these supergravities admit large global symmetry groups, noted by $G_D$. For example, maximal supergravities in $11-d$ dimension admit an $\Ed$ duality group \cite{Cremmer:1997ct}. Upon gauging a subgroup $G_g$ of $G_D$, this group becomes a \emph{duality group}, mapping a given gauged supergravity to an other (see \cite{Sezgin:2023hkc} for a survey of these supergravities). 

For supergravities, the gauging procedure is a bit more involved than for standard QFTs, as the gauging should be compatible with local supersymmetry. For example, one has to deal with the fact that the vectors of the ungauged supergravity transform in fixed representations $R_V$ of the duality group $G_D$. Thus, in order to define a covariant derivative one must introduce a constant ``embedding tensor", a linear map $\Theta:R_V \rightarrow \mathfrak{g}_D$ \cite{deWit:2002vt,deWit:2005hv,deWit:2005ub}. It allows us to define a covariant derivative:
\begin{equation}
    D_\mu = \nabla_\mu + A^A \Theta_A{}^\alpha t_\alpha\,
\end{equation}
where the index $A$ label the $R_V$ representation of the vector fields $A^A$ and $t_\alpha$ is a basis of the adjoint representation of $G_D$\footnote{In the following we will assume that $\mathfrak{g}_D$ is semi-simple such that we have a natural isomorphism $\mathfrak{g}^*_D \cong \mathfrak{g}_D$. We will also drop the distinction between $R_V$ and $R_V^*$.}. The embedding tensor must satisfy certain constraints imposed by supersymmetry and Jacobi identity. In particular, $\text{Im}(\Theta) = \mathfrak{g}_g$ must span a subalgebra of $\mathfrak{g}_D$ called the \emph{gauge algebra}. It must also satisfy a series of linear and quadratic conditions. If the embedding tensor satisfies those constraints, it is possible to improve the ungauged Lagrangian by adding terms in powers of $\Theta$ to make the ungauged action locally supersymmetric and invariant under $G_g$ gauged transformations. 

It was shown that the equations of motion of these gauged supergravities can be obtained from ExFT by specifying a \emph{reduction of the structure group} with singlet, constant intrinsic torsion, identified with the embedding tensor \cite{Cassani:2019vcl}. We will unpack these definitions in the rest of this section. However, we already note that they imply that the duality group $G_D$ must be the commutant of a compact group $G_S$ in $\Ed$:
\begin{equation}
G_D=\text{Comm}_{\Ed}\left(\,G_S\right)\,.
\end{equation}
In particular $G_D \subset \Es$. This prevents us from uplifting gauged supergravities with large field contents\footnote{In contrast with EFT techniques.}. The group $G_S$ is called the ``\emph{structure group}" of the truncation and the indices $A,\,B,\,C,\,...$ actually label $G_S$ singlets.

\paragraph{Reduction of the structure group}

For our purposes, a reduction of the structure group of the generalised tangent bundle is given by one of the two equivalent objects:
\begin{itemize}
    \item A globally well defined generalised frame $E\in G_S\backslash\Ed$ on $M_{\text{int}}$, or,
    \item A set of sections of the generalised tangent bundle, $K_A$, such that $\text{Stab}\langle K_A(p)\rangle =G_S(p) \cong G_S$.
\end{itemize}
An important remark is that, although the definition of the frame requires to select a specific group $G_S \in \Ed$, the sections specify a group $G_S(p)$ at each point $p \in M_{\text{int}}$. This group is conjugate to $G_S$ but $G_S(p)$ does not have to be the same set as $G_S$ seen as embedded in $\Ed$. Those two groups are related via the relation $G_S(p) := E(p)^{-1}\cdot G_S \cdot E(p)$.

In particular, we can build the invariant sections by using the projector $\mathbb{P}$ of $R_1$ on its $G_S$-invariant subspace:
\begin{equation}
    K_A = \mathbb{P}_A{}^M E_M\,.
\end{equation}
These invariant sections will be important to write down the truncation ansatz. 

\paragraph{Intrinsic torsion} Given a reduction of the structure group, one can build connections, $\nabla$, which are compatible with the reduction in the sense that:
\begin{equation}
    \nabla K_A = 0\,.
\end{equation}
One can then compute the torsion $T^\nabla$ of this connection, using the formula
\begin{equation}
    T^\nabla = L^\nabla - L
\end{equation}
This torsion is a tensor but does depend on the choice of compatible connection $\nabla$. Although compatible connections are not unique, given two compatible connections, $\nabla$ and $\nabla'$, their difference  $\nabla' - \nabla = \Omega$ can be identified with an element of  $E \otimes \mathfrak{g}_S$ (because $\mathfrak{g}_S$ leaves the section invariants). Thus, quotienting the torsion by the action of $E\otimes \mathfrak{g}_S$ on the compatible connections gives us a unique torsion tensor the ``\emph{intrinsic torsion}", independent of the choice of compatible connection $\nabla$. This object only depends on the specifics of the reduction of the structure group. The reduction of the structure group will define a consistent truncation if the intrinsic torsion is a constant $G_S$-singlet.

In term of the invariant sections, such an intrinsic torsion is equivalent to a tensor $X_{AB}{}^C$ defined as
\begin{equation}
    L_{K_A} K_B = -X_{AB}{}^C K_C\,.
    \label{eq:torsion1}
\end{equation}
The embedding tensor of the resulting truncated theory will be identified with the intrinsic torsion:
\begin{equation}
    X_{AB}{}^C = \Theta_A{}^\alpha (t_\alpha)_B{}^C\,.
    \label{eq:defXtheta}
\end{equation}

\paragraph{Truncation ansatz}
We can now expand the ExFT fields in terms of these invariant tensors coupled to the appropriate lower-dimensional objects. In particular, for the generalised metric and vectors, we get:
\begin{equation}
    \begin{split}
        &\mathcal{M}(x, Y) = E(Y)^T \cdot M(x) \cdot E(Y)\,,\\
        &\mathcal{A}_\mu{}^M (x,\,Y) = A_\mu{}^A(x) \, K_A{}^M(Y)\,,\\
    \end{split}
\end{equation}
where $M$ and $A_\mu$ are the lower-dimensional scalar matrix and vector fields, and $\mathcal{M}$ and $\mathcal{A}_\mu$ are the ExFT fields. Then, it ``suffices" to use the ExFT dictionary to obtain the consistent truncation expressed in terms of supergravity fields. This includes performing a series of non-linear field redefinitions which are compiled in appendix \ref{app:KKCalculus} for the $\Es$-ExFT/type IIB case.

\section{Construction of an uplift}
\label{sec:Construction}

In this section, we will construct the uplifts of a gauged supergravity theory in $D$ dimensions to a higher-dimensional 10D/11D supergravity, given the following data:
\begin{itemize}
    \item The duality group of the lower-dimensional theory, denoted by $G_D = \text{Comm}_{G_S}(\Ed)$, where $G_S \subset \Ed$ is the structure group;
    \item The embedding tensor of the reduced theory, $\Theta_A{}^\alpha$ (or equivalently $X_{AB}{}^C$).
\end{itemize}
We assume that the reduced theory has a non-trivial gauge group, ensuring the presence of invariant 1-forms $A^A$ and associated sections $K_A$ of the generalised tangent bundle with constant singlet intrinsic torsion. In particular we will show that the internal manifold can be modelled as
\begin{equation}
    \Mint = G_g/H\times B\,.
\end{equation}
where $B$ is topologically trivial. Then, we will show that the consistency of the truncation can be inferred from a series of PDE not on the full $\Mint$, as in \eqref{eq:torsion1}, but on $B$ only.

\subsection{The internal manifold}
Assume we have a set of non-vanishing generalised sections $K_A$ with constant singlet intrinsic torsion. We first examine the vector components:
\begin{equation}
    k_A := K_A^{(1,\,0)} \in T\Mint\,.
\end{equation}
Note that, on the vector subbundle $T\Mint \subset E$, the generalised Lie derivative reduces to the standard Lie derivative. The torsion condition becomes:
\begin{equation}
    (L_{K_A} K_B)^{(1,0)}= \mathcal{L}_{k_A} k_B =-X_{AB}{}^C k_C \,.
    \label{eq:TorsionVectors}
\end{equation}
This directly implies that the vector fields $k_A$ close as a real Lie algebra $\mathfrak{g}_g$, the gauge algebra, a subalgebra of vector fields algebra denoted $\mathfrak{X}(\Mint)$. Choosing a basis of $\mathfrak{g}_g \in \mathfrak{X}(\Mint)$, $k_\alpha$, we can write 
\begin{equation}
    k_A = \Theta_A{}^\alpha k_\alpha\,.
\end{equation}
We have implicitly assumed that the action of $\mathfrak{g}_g$ on $\Mint$ is \emph{effective}. If not, solving the full torsion constraint will show that some sections $K_A$ vanish\footnote{Note that this argument does not hold for the modified generalised Lie derivative used to describe the massive IIA supergravity equations of motion as built in \cite{Ciceri:2016dmd}}.

\subsubsection{A local model for $\Mint$}
We will assume that the action of this Lie algebra lifts to a \emph{proper action} for a Lie group $G_g$ associated to the algebra $\mathfrak{g}_g$. This assumption, which is always valid for compact Lie groups, allows us to use the results of \cite{Meinrenken2003}, reviewed in appendix \ref{app:GroupActions}, and give some structure to $\Mint$. Importantly for us, it will allow us to study quotients $\Mint/G_g$ avoiding degenerate (non-Hausdorff) quotient spaces\footnote{We aim at avoiding situations such as the quotient of $T^2$ by $\mathbb{R}$ acting as $(x,\,y) \rightarrow (x+ t,\,y+\alpha \,t)$ with $\alpha \notin \mathbb{Q}$.}. 

In particular, we can show that there exists a distinguished compact subgroup of $G_g$  called the ``\emph{principal stabiliser}" $H_{\text{prin}}$ such that a generic point on $\Mint$ is stabilised by a group conjugate to $\Hprin$.  More precisely:
\begin{itemize}
    \item $\Hprin$ is the ``smallest stabiliser" on $\Mint$. If $H_p$ is the stabiliser of a point $p \in \Mint$ then $(\Hprin) \leq (H_p)$, i.e. up to $G_D$-conjugation $H_{\text{prin}} \subset H_p$.
    \item Almost all points are stabilised by $\Hprin$. Let $M_{(H_{\text{prin}})} = \{p \in \Mint\,|\, (H_{p}) = (H_{\text{prin}})\}$ be the set of points $p$ whose stabilisers under the action of $G_g$ are conjugate to $H_{\text{prin}}$, then $M_{(\Hprin)}$ is an open dense smooth embedded submanifold of $\Mint$.
    \item The space $\Mint$ can be decomposed in ``strata" $M_{(H)}$ labelled by conjugacy classes of subgroups of $G_g$. Each of these strata are smooth manifolds and so are their quotients $X_{(H)}=M_{(H)}/G_g$. Moreover, the quotient map is a submersion and any $X_{(H)}$ is an embedded submanifold of $X=\Mint/G_g$ for $H$ a subgroup of $G_g$.
    \item The principal stratum $X_{(\Hprin)} = M_{(\Hprin)}/G_g$ is open and dense in $X=\Mint/G_g$.
\end{itemize}
These results give us a local model for the internal manifold. In the neighbourhood of any point in $M_{(\Hprin)}$ we can write
\begin{equation}
    M_{\text{loc}} =  G_g/\Hprin \times B \,,
\end{equation}
where $B$ will be referred to as the \emph{base} of the internal space and is topologically trivial, whereas $G_g/\Hprin$ will be called the \emph{fibre} of the internal space. In this model, the base $B$ is a local trivialisation of $X_{(H_{\text{prin}})}$. This model is very useful since $G_g$ leaves the base invariant while it acts transitively on the fibre. Since a consistent truncation is a statement about local equations of motion, it is reasonable to reduce the problem to a consistent truncation on $M_{(\Hprin)}$ and then to a consistent truncation on $\Mloc$. Of course, one should be careful when extending these results to $M_{(\Hprin)}$ then to the full of $\Mint$, where topological obstructions might arise. In the following, we will drop the ``prin'' subscript of $\Hprin$. Moreover, we will make the distinction between $H$ seen an abstract group isomorphic to $\Hprin$, and the specific group $H_D$ embedded in $G_g \subset G_D \subset \Ed$ and isomorphic to $H$. 
 
 When using local coordinates, we will use the following notations:
\begin{equation}
\begin{array}{l}
    \underbrace{y^m}_{\text{coord on } M_{\text{loc}}}\rightarrow \underbrace{y^{\tilde{m}}}_{\text{coord on the base } B} \oplus \underbrace{y^{\bar{m}}}_{\text{coord. on the fibre } G/H}\,. 
\end{array}
\end{equation}
When a vielbein is defined, we will use the same notations but underlined for flat indices $({\underline{m}} \,\rightarrow \,{\underline{\tilde{m}}}\, \oplus\, {\underline{\bar{m}}} $).

\subsubsection{Homogeneous spaces}

Since the fibre is an homogeneous space, we recall here some facts concerning their geometry. Consider the homogeneous spaces $G/H$ for $G$ a Lie group and $H$ a compact subgroup of $G$. Given a choice of coset representative $L(p) \in G$ for any point $p\in G/H$, we would like to build a vielbein and a basis for the vectors of $\mathfrak{g} \subset \mathfrak{X}(G/H)$. When acting with $g\in G$ from the left, the coset representative changes as
\begin{equation}
    g\cdot L(p) = L(g\cdot p) \cdot h(p,\,g)\,.
    \label{eq:cosetRepTransformations}
\end{equation}
where $h(p,\,g) \in H$ is called the ``compensator". We define the Maurer-Cartan form $\Omega$:
\begin{equation}
    \Omega = L^{-1} dL \in \Omega^1(\mathfrak{g})\,.
\end{equation}
This is a 1-form with value in $\mathfrak{g}$.
 We can orthogonally split 
\begin{equation}
    \mathfrak{g} = \mathfrak{K} \oplus \mathfrak{h}
\end{equation} 
by using a $G$-invariant metric on $\mathfrak{g}$, for example the Cartan-Killing metric if it is non-degenerate. Separated in this way, the Maurer-Cartan form reads
\begin{equation}
    \Omega = \underbrace{\mathring{e}}_{\in \mathfrak{K}} + \underbrace{Q}_{\in \mathfrak{h}}\,.
    \label{eq:vielbeinGH}
\end{equation}
This definition depends  on the choice of coset representative $L(p)$. Let us see how they transform upon choosing another representative $L'(p) = L(p)\cdot h(p)$. A quick computation yields that
\begin{equation}
    \Omega' = L'^{-1} dL' = h^{-1}\cdot \Omega \cdot h + h^{-1} dh\,.
\end{equation}
This shows that the flat index of the vielbein transforms under the adjoint representation,\footnote{Choosing a basis $t_{\underline{\bar{m}}}$ on $\mathfrak{K}$ and a basis $y^{\bar{m}}$ on $G/H$, it reads $\mathring{e}_{\bar{m}}{}^{\underline{\bar{m}}}$ and is indeed a 1-form with value in $\mathfrak{K}$.} while $Q$ does transform as an $H$-connection for the $H$-principal bundle $G\rightarrow G/H$.

In the same way, the action of $G$ by diffeomorphisms transforms the Maurer-Cartan form as
\begin{equation}
\begin{array}{ll}
   (L_{g^{-1}}^* \Omega)(p) &= (L^{-1}(g\cdot p)) d(L(g\cdot p ))_{|p} \\ 
   &=(g^{-1}\cdot L(p)\cdot h(p,\,g))^{-1}d(g^{-1}\cdot L(p)\cdot h(p,\,g))_{|p}\\
   &=h^{-1}(p,\,g)\cdot \Omega(p) \cdot h(p,\,g) + h^{-1}(p,\,g)d( h(p,\,g))\,.
    \end{array}
    \label{eq:changeVielbeinH}
\end{equation}
From this we obtain that $(L_{g^{-1}}^*\mathring{e})(p) =  \text{Adj}(h)\cdot \mathring{e}(p)$. This was expected from the fact that the structure group on $G/H$ is $H$ itself.

\subsubsection{Killing vector and the reduced embedding tensor} 
The vector fields $k_A$ on the homogeneous space $G_g/H$, can be rewritten as
\begin{equation}
    \Theta_A{}^\alpha k_\alpha = L_A{}^B \Theta_B{}^{\underline{\bar{m}}} (\mathring{e}^{-1})_{\underline{\bar{m}}} 
    \label{eq:KillingSections}
\end{equation}
where $L \in G_g$ is a choice of coset representative in the appropriate representation, and $\mathring{e}$ is the associated vielbein. This defines a reduced embedding tensor $\Theta_A{}^{\underline{\bar{m}}}$ where the index $\underline{\bar{m}}$ should be thought of as a flat index on $G_g/H$. Note that the reduced embedding tensor depends on the choice of principal stabiliser $H$, up to conjugation. 

Finally, putting all the pieces together, we obtain that the vector piece of the generalised sections can be written as
\begin{equation}
    (K_A)^{(1,\,0)} = \Theta_A{}^\alpha k_\alpha =L_A{}^B \Theta_B{}^{\underline{\bar{m}}} (\mathring{e}^{-1})_{\underline{\bar{m}}}{} \in \Gamma(TM_{(H)})\,,
    \label{eq:vecPartSections}
\end{equation}
where we have used the existence of smooth embeddings of the orbits $\phi_p:G_g\cdot p \rightarrow TM_{(H)}$ $\forall p \in \Mint$ to embed $T(G_g/H)$ in $TM_{(H)}$.

\subsection{The generalised sections}
\label{sec:ClosedFroms}
In this section we will show that, upon making some assumptions on the topology of $G_g/H$ and the compactness of the gauge group, the $p$-form components of the generalised sections are closed while the $p$-vector densities are divergenceless (i.e. their duals are closed). This will dramatically simplify the constraints obtained from requiring the torsion to be a constant $G_S$-singlet.

\subsubsection{Constraints from trivial gauge transformations}

Let us consider the would-be uplift of the origin of the scalar manifold of the lower-dimensional supergravity, setting to zero vector fields and $p$-forms. Consider the $G_S$-invariant sections $\tilde{K} \in \text{Ker}(\Theta)$\footnote{Here we consider the embedding tensor as a linear map from the vector space of the sections $\langle K_A\rangle$ to the gauge algebra $\mathfrak{g}_g$.}, i.e. the sections such that $\tilde{K}^{(1,\,0)} = 0$. The infinitesimal action associated to these sections should be trivial, since they correspond, in the lower-dimensional supergravity, to spurious gauge transformations. In particular, the action on the $p$-form potentials $C_{(p)}$ is given by
\begin{equation}
    L_{\tilde{K}} C_{(p)} = \underbrace{\mathcal{L}_{\tilde{K}^{(1,\,0)}} C_{(p)}}_{=0} + \,\text{gauge transformations} \stackrel{!}{=} 0\,.
    \label{eq:genLGaugeTr}
\end{equation}
Unpacking the gauge transformations term by term, both in the type IIB case and in the M-theory case for the $\Es$-ExFT, we obtain that
\begin{equation}
    \begin{array}{l|l}
    \text{Type IIB} & \text{M-theory}\\[2mm]\hline
       \delta_{\tilde{K}} \mathbb{B}_2^\alpha  = d {\tilde{K}}^{(0,\,1)\,\alpha} \phantom{\Big{(}} &  \delta_{\tilde{K}} A_3 = d{\tilde{K}}^{(0,\,2)}\\[2mm]
         \delta_{\tilde{K}} C_4 = d {\tilde{K}}^{(0,\,3)} +\, \tfrac{1}{2} \epsilon_{\alpha\beta} d {\tilde{K}}^{(0,\,1)\,\alpha}\wedge \mathbb{B}^\beta_2 \phantom{\Big{(}} &  \delta_{\tilde{K}} A_6 = d{\smallstar\tilde{K}}^{(2,\,0)} + d{\tilde{K}}^{(0,\,2)} \wedge A_3\\[2mm]
        \epsilon_{\alpha\beta} \delta_{\tilde{K}} \mathbb{B}_6^\beta = d\left(\smallstar {\tilde{K}}^{(1,\,0)}_{+1\,\alpha}\right) + (\text{terms} \propto d{\tilde{K}}^{(0,\,1)\,\alpha}) \phantom{\Big{(}} &\\ 
    \end{array}
\end{equation}
By equation \eqref{eq:genLGaugeTr}, these transformations must vanish. It implies that 
\begin{equation}
    \begin{array}{l|l}
    \text{Type IIB} & \text{M-theory}\\[2mm]\hline
      \hspace{5mm}  d\tilde{K}^{(0,\,1)} =  0 \phantom{\Big{(}}\hspace{1cm} &\hspace{5mm} d\tilde{K}^{(0,\,2)} = 0\phantom{\Big{(}} \\
     \hspace{5mm}   d\tilde{K}^{(0,\,3)} = 0 \phantom{\Big{(}} &\hspace{5mm}  d \smallstar \tilde{K}^{(2,\,0)}\\
    \hspace{5mm}  d\smallstar \tilde{K}^{(1,\,0)} = 0 & 
    \end{array}
\end{equation}
which shows that the corresponding sections are built out of (co-)closed $p$-forms. This result generalises trivially to other $\Ed$-ExFTs with $d\leq 7$.

\subsubsection{Constraints from compact gauge transformations}
The same statement can be extended to the sections $\mathring{K} \in \Theta^{-1}(\mathfrak{K})$ where $\mathfrak{K}$ is the Lie algebra associated to the maximally compact subgroup $K\subset G_g$ (note that $\text{Ker}(\Theta) \subset \Theta^{-1}(\mathfrak{K})$). These are the sections whose vector component corresponds to a ``compact" Killing vector. Again, we consider the uplift of the lower-dimensional theory, setting all fields to zero. This configuration is invariant under the maximally compact subgroup $K\subset G_g$ and so is its uplift. This means  that both the metric and $p$-form field strengths $F_{(p)}$ are $K$-invariant. An averaging argument shows that, when the fluxes are exact, the $(p-1)$-form potential themselves can be chosen to be invariant under $K$
\footnote{
Indeed, choosing a $(p-1)$-form potential $\tilde{C}_{(p-1)} \in \Omega^{p-1}(\Mint)$ such that $F_{(p)} = d \tilde{C}_{(p-1)}$. We define
\begin{equation}
    C_{(p-1)} = \frac{1}{|K|} \int_K dk \,\left(L_k^*\circ \tilde{C}_{(p-1)} \right)\,,
    \label{eq:AverArgInv}
\end{equation}
where we used the Haar measure on $K$, giving a finite volume $|K|$ to $K$. With these definitions, $d C_{(p-1)} = F_{(p)}$ yielding a $K$-invariant potential.}. This implies that there exists a gauge where both
\begin{equation}
    \delta_{\mathring{K}} C_{(p)} = 0 \hspace{1cm}\text{and}\hspace{1cm} \mathcal{L}_{\mathring{k}} C_{(p)} = 0\,.
\end{equation}
This reduces us to the case considered in \eqref{eq:genLGaugeTr}, implying that the generalised vectors $\mathring{K}$ are built out of (co-)closed $p$-forms. Equivalently, this whole procedure can be thought of as performing a partial gauge fixing for the $p$-form potentials.

\subsubsection{The topological conditions}

We have shown that any section $\mathring{K}_A \in \Theta^{-1}(\mathfrak{K})$ is built out of (co-)closed forms on $\Mint$ if the most generic closed $K$-invariant $p$-form fluxes are exact. This always happen when $H^p(\Mint)=0$ for $p=1,\,3,\,5$ and $7$ in type IIB, or $p=4$ and $7$ in M-theory. Since the $G_g$-action closes on $\Mloc$ and $H^q(B)=0$ for $q\geq 1$, by K\"unneth formula, it is sufficient to impose the weaker constraint
\begin{equation}
    H^p(G_g/H) = 0 \,\,\text{ for }\,\,\begin{cases} p =1,\, 3,\,5 \text{ and } 7 \text{ in type IIB,}\\[2mm]
    p = 4 \text{ and } 7 \text{ in M-theory.}\end{cases}
    \label{eq:topConditions}
\end{equation}
We insists that to obtain these simplifications almost everywhere on $\Mint$, we do not require any constraints on the full $\Mint$ but only on $G_g/H$. 

When these topological conditions are met, the previous observations show that the torsion constraint reduces to 
\begin{equation}
    L_{\mathring{K}_A} \mathring{K}_B = \mathcal{L}_{\mathring{k}_A} \mathring{K}_B =- X_{AB}{}^C \mathring{K}_C\,.
    \label{eq:RedTorsionConstraints}
\end{equation}
As such the torsion constraint simply signals that the $\mathring{K}_A$ are built out of \emph{equivariant} and (co-)\emph{closed} $p$-forms.

In appendix \ref{app:Equiv}, we show that on $G_g/H$, equivariant $p$-forms with value in a representation $V$ are in one-to-one correspondence with $H$-invariant tensors of $\Lambda^p T^*_e(G_g/H) \otimes V$. Since the same argument holds for each independent orbit, a $V$-valued equivariant form $\omega$ can be entirely specified by a function $\tilde{\omega} : B \rightarrow \Lambda^pT^*_e \Mloc \otimes V$ which is smooth and $H$-invariant. Obviously, one needs to be careful when patching different strata.

\subsection{The generalised frame}
\label{subsec:GenFrame}

Studying the sections and the torsion conditions, we have shown that the sections must be built out of equivariant and (co-)closed $p$-forms. We also identified the vector component of the sections with the Killing vectors on $G_g/H$. One still needs to impose that $\text{Stab}(K_A) = G_S$. This is more easily done by studying the frame corresponding to these sections. 

\subsubsection{Generalised frames in ExFT} 

A generalised frame, in the ExFT context, is a section of $K_d\backslash \Ed$ where $K_d$ is the maximally compact subgroup of $\Ed$. We will first discuss how to embed the frame $\mathring{e}$ on $G_g/H$ in $\Ed$. This is done by specifying a solution to the section constraint \eqref{eq:secConstraints}. Such a solution is given by a tensor $\mathcal{E}_M{}^m$ transforming from the left by $\EdR$ elements in the $R_1$ representation, and on the right by a $\GL(d')$ geometric group\footnote{$d' = d$ or $d-1$ depending on the type of embedding (in type IIB or in M-theory).} This defines uniquely an embedding 
\begin{equation}
\GL(d')\rightarrow \GL(d')_\mathcal{E} \subset \Ed\times \mathbb{R}^+
\end{equation}
via the relation
\begin{equation}
    g_M{}^N \mathcal{E}_N{}^n = \mathcal{E}_N{}^m g_{m}{}^n\hspace{1cm}\text{where}\hspace{1cm} g_M{}^N \in \Ed\times \mathbb{R}^+ \text{ and } g_{m}{}^n\in \GL(d')\,.
\end{equation}
The subscript $\mathcal{E}$ reminds us that the specific embedding might depend on the choice of solution to the section constraint $\mathcal{E}$. Indeed, for a given class of solutions to the section constraint all the embeddings of $\GL(d')_\mathcal{E}$ in $\EdR$ are conjugate to one another but not strictly identical. Embedding $\mathring{e}$ in $\GL(d')$ and then in $\GL(d')_\mathcal{E}$ we obtain its embedding in $\Ed$.

We also define another relevant subgroup of $\EdR$, the stabiliser of $\mathcal{E}$:
\begin{equation}
    \mathcal{S}_\mathcal{E} = \{S \in \EdR\,|\,S\cdot \mathcal{E} = \mathcal{E}\}\,.
\end{equation}
Using the supergravity dictionary, this group encodes the contributions of the fluxes to the generalised frame (and that of the axio-dilaton matrix in the type IIB case). This group also contains the trombone symmetry of maximal supergravities. The action by conjugation of $\GL(d')_\mathcal{E}$ closes on $\mathcal{S}_\mathcal{E}$.

\subsubsection{Compatibility constraints} 

The frame defining a reduction of the structure group is a global \emph{non-vanishing} section of a $G_S\backslash \left(\EdR\right)$ bundle on $\Mint$. Using the torsion constraint, one can show that this frame defines the same reduction of the structure group, with the same torsion, as the sections $K_A$ if
\begin{equation}
    K_A{}^N = \mathbb{P}_A{}^M \, E_M{}^N\,.
    \label{eq:Compconstraints}
\end{equation}
Here $\mathbb{P}$ is the projector on the $G_S$-invariant subspace of the representation $R_1$. The existence of a frame satisfying \eqref{eq:Compconstraints} implies that $\text{Stab}(K_A)$ is precisely congruent to $G_S$ at any point in $\Mint$. This condition is called the ``\emph{compatibility constraint}". 

We can now describe the most generic frame on $\Mloc$ satisfying the compatibility constraint \eqref{eq:Compconstraints} for sections satisfying \eqref{eq:vecPartSections}. The torsion constraint implies that
\begin{equation}
   ( \mathbb{P} E\mathcal{E})_A{}^m= K_A{}^m = \Theta_A{}^\alpha k_\alpha{}^m = L{}_A{}^B \Theta_B{}^{\underline{m}} (\mathring{e}^{-1})_{\underline{m}}{}^m\,.
   \label{eq:vectConstraints}
\end{equation}
We factor out $L$ and $\mathring{e}$ which defines a flattened frame $E^\flat$
\begin{equation}
     E(p) = L(p) \,E^\flat(p)\, \mathring{e}^{-1}(p)\,.
\end{equation}
The equation \eqref{eq:vectConstraints} imposes that 
\begin{equation}
\mathbb{P}_A{}^M\,E^\flat_M{}^N(p) \,\mathcal{E}_N{}^{\underline{m}} = \Theta_A{}^{\underline{m}}\,,
    \label{eq:Eflatcondition}
\end{equation} 
which prompts the definition of the set
\begin{equation}
    E_{\text{comp}}^{\mathcal{E}} = \{ E \in \EdR\,|\, (\mathbb{P}\cdot E\cdot \mathcal{E})_M{}^m = \Theta_M{}^m\}\,.
    \label{eq:EcompSimp}
\end{equation}
This set contains all admissible flat frames but is not a group. To provide a simpler ansatz, we need to characterise further this set. Given a single element $E_0 \in E_{\text{comp}}^\mathcal{E}$, any element of $E_{\text{comp}}$ can be decomposed as
\begin{equation}
E_{\text{comp}}^{\mathcal{E}} = G_S \cdot \text{Stab}(\mathcal{E}_{\text{comp}})\cdot E_0 \cdot \mathcal{S}_\mathcal{E}\,.
\label{eq:ECompDec}
\end{equation}
In this decomposition, we have defined the space of compatible solutions to section constraint $\mathcal{E}_{\text{comp}}$ as
\begin{equation}
    \mathcal{E}_{\text{comp}} =\left\{\mathcal{E}_M{}^{\underline{m}} \,| \,Y^{MN}{}_{PQ} \mathcal{E}_M{}^{\underline{m}} \mathcal{E}_{N}{}^{\underline{n}}=0\hspace{2mm}\text{and}\hspace{2mm}\mathbb{P}_A{}^M\mathcal{E}_M{}^{\underline{m}} = \Theta_A{}^{\underline{m}} \right\}\,,
    \label{eq:compSectionConstraintSet}
\end{equation}
and $\text{Stab}(\mathcal{E}_{\text{comp}})$ is the stabiliser of the $\mathcal{E}_{\text{comp}}$ vector space. As such, selecting a $E_0$ in $E_{\text{comp}}$, $\mathcal{E}_0 = E_0\cdot \mathcal{E}$ is an element of $\mathcal{E}_{\text{comp}}$ which explains the decomposition \eqref{eq:ECompDec}.

The flat frame $E^\flat$ can be further constrained. Assuming the topological constraint \eqref{eq:topConditions}, $E^\flat$ can be chosen to be $H_\mathcal{E}$-invariant. The group $H_{\mathcal{E}}$ is defined as the subgroup of $\GL(d')_{\mathcal{E}}\subset \EdR$ specified by the transformation rules of the vielbein under a change of coset representative:
\begin{equation}
    \mathring{e} \rightarrow h \cdot \mathring{e}\,.
\end{equation}
By invariance we mean that $h \cdot E^\flat \cdot h^{-1} = E^\flat$. This allows us to further reduce the study of $E^{\mathcal{E}}_{\text{comp}}$ to its $H_{\mathcal{E}}$-invariant subset. Moreover, if there exists a compatible solution to the section constraint then $\forall h_\mathcal{E} \in H_\mathcal{E}$, $\exists \,h_D \in H_D$ and $h_S \in G_S$ such that 
\begin{equation}
    h_\mathcal{E} = h_D \cdot h_S\,.
\end{equation}
This decomposition is unique up to conjugation and defines a group $H_S \subset G_S$ isomorphic to a subgroup of $H$. This property will prove useful when working out examples. If furthermore we assume $G_g$ to be compact, one can show that we can choose a representative $E^\flat$ which is constant on $G_g$-orbits. Proofs can be found in appendix \ref{app:EflatConstraints}.

We note that it is often easier to characterise the set $\mathcal{E}_{\text{comp}}$ than the set $E^\mathcal{E}_{\text{comp}}$. Importantly, if $\mathcal{E}_{\text{comp}}$ is empty, there is no uplift with the specific choice of local model $\Mloc$ and principal stratum $H$. We also remark that the existence of a solution to the section constraint and their type (IIB/M-theory) depends on the choice of principal stabiliser $H$ and not on $G_g$. Furthermore, for each choice of solution to the section constraint there exists at most one type of principal orbit type $H$ and one group $H_S$.

In conclusion, we have reduced the space of possible generalised frames to frames in \eqref{eq:EcompSimp}, constant on $G_g$-orbit (i.e. which only depend on the coordinates on the base) and $H_\mathcal{E}$-invariants. The sections associated to such frames are, by construction, equivariant and satisfy the compatibility conditions. Finally, the torsion constraint reduces to 
\begin{equation}
d\left[\mathbb{P}\left(L\cdot E^\flat \cdot \mathring{e}^{-1}\right)_{(p)}\right] = 0\,,
\end{equation}
where $d$ denotes the exterior derivative (or its $\smallstar$-dual) acting on the $p$-form (or on vector densities) obtained by splitting the generalised vectors in poly-forms. Factoring out $L$ and $\mathring{e}$ these equations reduce to PDEs on $B$ only.

\subsection{Summary of the construction}
We summarise both our ansatzes in terms of sections $K_A$ and in terms of the frame $E$. We will assume from now on that $G_g$ is compact and that the topological conditions \eqref{eq:topConditions} are satisfied. We start by summarising the different notations for the relevant groups and their embeddings in $\EdR$.
\begin{itemize}
	\item $G_S$ is the structure group. It is constant and defined from the beginning by the field content of the lower-dimensional supergravity we want to uplift.
	\item $G_D$ is the duality group. It is the commutant of $G_S$ in $\Ed$.
        \item $G_g \subset G_D$ is the gauge group of the lower-dimensional supergravity.
	\item $H \subset G_g$ is the isotropy group of a generic point in $\Mint$. Several choices are possible and impact the possibility of an uplift in type IIB/M-theory. We distinguish this abstract group from its embedding $H_D$ in $G_D \subset\Ed$.
	\item $\GL(d')_\mathcal{E}$ is the geometric group embedded in $\Ed$ by using a solution to the section constraint $\mathcal{E}$. Its precise embedding depends on $\mathcal{E}$. Its rank is $d'=d$ if we choose a M-theory solution to the section constraint, and $d'=d-1$ if we choose the type IIB solution to the section constraint.
	\item $H_{\mathcal{E}} \subset \GL(d')$ is the embedding of $H$ in $\GL(d')_{\mathcal{E}}$. When there exists an uplift, there exists a specific $\mathcal{E}$ such that $\forall\,h_\mathcal{E}  \in H_{\mathcal{E}}$, $\exists\,h_D \in H_D$ and $\exists h_S \in G_S$ such that $h_\mathcal{E} = h_D \cdot h_S$.
\end{itemize}
We build consistent truncations in three steps. In step one we classify ``compatible" solutions to the section constraint. In step two, we provide the most generic ansatz for a frame satisfying the compatibility constraint. Finally, in step three, one imposes a differential condition on the frame in order to solve the torsion constraint. This differential condition reduces to a set of PDEs on the base $B$.

\paragraph{Compatible solutions to the section constraint} 

To build the consistent truncation, one starts by classifying the possible principal stabilisers $H\subset G_g$, up to conjugation. For an uplift to exist, we require that there exists a solution to the section constraint $\mathcal{E}_0$ such that
\begin{equation}
(\mathbb{P}\mathcal{E}_0)_A{}^{\underline{m}} = \Theta_A{}^{\underline{m}}\,.
\end{equation}
Since solving explicitly the section condition $Y^{MN}{}_{PQ} \mathcal{E}_M{}^m \mathcal{E}_N{}^n = 0$ might be computationally expensive, we have to consider the simplest possible ansatz for $\mathcal{E}_0$. Obviously 
\begin{equation}
    (\mathcal{E}_{0})_{A} = \Theta_A\,.
\end{equation}
However, for the non $G_S$-singlet of $\mathcal{E}_0$, we can simplify the ansatz by trying inequivalent guesses for the embeddings of $H_{\mathcal{E}}$ in $H_D \times G_S$ and only solving the section constraint for the $H_{\mathcal{E}}$-invariant sections. Once a solution is found, it specifies the unique possible embedding of $H_\mathcal{E}$ in $H_D \times G_S$ compatible with the existence of a consistent truncation. This also specifies the type of the uplift, in type IIB or M-theory. If there is no solution for any of the inequivalent embeddings of $H \rightarrow G_D \times G_S$, then there are no uplifts with the given choice of principal stabiliser $H$. 

\paragraph{The frame ansatz} The previous computation will specify the set
\begin{equation}
\mathcal{E}_{\text{comp}}^H = \{\mathcal{E}_M{}^m \, |\, h_\mathcal{E}\cdot \mathcal{E}\cdot h^{-1} = \mathcal{E},\,\mathcal{E}_A{}^{\underline{m}} = \Theta_A{}^{\underline{m}},\,Y^{mn}{}_{PQ} = 0\}\,.
\end{equation}
From it, one can work out the $H$-invariant elements of $\text{Stab}(\mathcal{E}^H_{\text{comp}})$ (as well as optionally an element $E_0\in \Ed$ sending your favourite solution to the section constraint to an element in $\mathcal{E}_{\text{comp}}^H$). Moreover, the choice of principal stabiliser forces us to work on the manifold
\begin{equation}
    \Mloc =  G_g/H \times B\,,
\end{equation}
and the generalised frame on this manifold will be of the form 
\begin{equation}
    E = L\cdot E^\flat \cdot (\mathring{e}^{-1})\,.
    \label{eq:GenFrameAnsatz}
\end{equation}
We have denoted $L \in G_g$ a coset representative of $G_g/H$ and $\mathring{e}$ is the associated inverse vielbein. The flat frame $E^\flat$ can be written as the product 
\begin{equation}
    E^\flat = Z\cdot  E_0 \cdot S\,,
\end{equation}
for elements $Z \in \text{Stab}(\mathcal{E}^H_{\text{comp}})$ and $S \in \mathcal{S}_{\mathcal{E}}$. Both $Z$ and $S$ must be $H$-invariant functions of the basis $B$. This yields the most generic ansatz satisfying the compatibility condition and possibly the torsion conditions. The fact that the frame is globally well defined follows from the $H$-invariance of the flattened frame and its independence on the fibre coordinates.

\paragraph{The differential condition}

Finally, we get a differential condition on $E^\flat$, which only depends on the base $B$ by imposing that
\begin{equation}
    K_A{}^N = \mathbb{P}_A{}^{\underline{M}}E_{\underline{M}}{}^N\,.
\end{equation}
Since the section are built from (co-)closed forms, we have the constraint that
\begin{equation}
d\left[\mathbb{P}\cdot\left(L\cdot E^\flat \cdot \mathring{e}^{-1}\right)_{(p)}\right] = 0\,,
\label{eq:TorsionConstraintsFinal}
\end{equation}
i.e. each $p$-form components of the 
$K_A$ must be (co-)closed. Factoring out $L$ and flattening indices using $\mathring{e}$, this condition reduces to a set of PDE on the base. 

\section{M-theory embedding of \texorpdfstring{$\mathcal{N}=4$}{N=4} supergravity with \texorpdfstring{$n_v=6$}{nv=6}}
\label{sec:nv6Example}

We start by illustrating the above method with a somewhat trivial example. We consider the $\mathbb{Z}_2$-structure giving rise to a consistent truncation to $\mathcal{N}=4$ supergravity coupled to $6$ vector multiplets, admitting an $\SO(4)_L \times \SO(4)_R$ gauging and a supersymmetric AdS$_4$ vacuum. The duality group of this theory is $G_D = \SO(6,\,6) \times \SL(2)$. The maximal compact subgroup of $\SO(6,\,6)$ is $\SO(6)_L \times \SO(6)_R$. Here we consider the gauge group $G_g =\SO(4)_L \times \SO(4)_R$ where each $\SO(4)_{L,\,R} \subset \SU(4)_{L,\,R} \cong\SO(6)_{L,\,R}$. The gauging is purely electrical (i.e. no ``dyonic" gauging). We will show that the uplift of this theory is unique. It is a $\mathbb{Z}_2$-subsector of M-theory compactified on $S^7$ \cite{deWit:1982bul,deWit:1986oxb,Varela:2015ywx}.

To present this truncation, we will use the $\SL(8)$ decomposition of $\Es$. The fundamental and adjoint representations of $\Es$ branch as
\begin{equation}
\begin{split}
\mathbf{56}\rightarrow \mathbf{28}\oplus \mathbf{28}'\,,\\
\mathbf{133}\rightarrow \mathbf{63} \oplus \mathbf{70}\,.
\end{split}
\end{equation}
The explicit form of the $\es$ generators we use can be found in the appendix of \cite{Sterckx:2024vju}. The fundamental representation of $\SL(8)$ will be labelled by an index $\Lambda,\,\Sigma = 1,\,\dots,\,8$. The M-theory solution to the section constraint is given by writing
\begin{equation}
\mathcal{E}_{[m8]}{}^n = \delta_m^n \hspace{1cm} m,\,n=1,\,\dots,\,7\,.
\end{equation}
with all other entries vanishing. The structure group $G_S = \mathbb{Z}_2$ is generated by the $\SL(8)$ element
\begin{equation}
\sigma = \begin{pmatrix}
	\mathbb{1}_{4\times 4} & 0 \\
	0 & -\mathbb{1}_{4\times 4}
\end{pmatrix} \in \SL(8) \subset \Es\,.
\end{equation}
Embedded in $\Es$, the $\SO(4)_{L,\,R}$ groups act on the first or last four indices of the fundamental $\SL(8)$ representation. As principal stabiliser, we will select 
\begin{equation}
    H = \SO(3)_L \times \SO(3)_R
\end{equation} acting on the indices $2,\,3,\,4$ and $6,\,7,\,8$ respectively.

We start by specifying a coset representative for the $G_g/H = S^3 \times S^3$ manifold: 
\begin{equation}
\begin{array}{rl}
L = &\exp\left(-(\phi_3-\pi/2) R_{[34]}\right)\cdot\exp(-\phi_2 R_{[23]})\cdot\exp(-\phi_1 R_{[12]})\\
&\cdot\exp(-(\theta_3 - \pi/2) R_{[78]})\cdot\exp(-\theta_2 R_{[67]})\cdot\exp(-\theta_1 R_{[56]})\,,
\end{array}
\end{equation}
where $R_{[\Lambda\Sigma]}$ is the $\SL(8)$ rotation in the plane spanned by $\Lambda$ and $\Sigma$. We have introduced angular coordinates $\theta_i$ and $\phi_i$ on each of the two three-spheres. Using \eqref{eq:vielbeinGH}, this specifies a vielbein 
\begin{equation}
\begin{array}{rl}
\mathring{e} =& d\phi_1 + \sin(\phi_1) d\phi_2 + \sin(\phi_1) \sin(\phi_2) d\phi_3\\
&+ d\theta_1 + \sin(\theta_1) d\theta_2 + \sin(\theta_1) \sin(\theta_2) d\theta_3\\
&+ e(\alpha) d\alpha\,.
\end{array}
\end{equation}
Of course, the vielbein along the base, with coordinate $\alpha$, is unspecified. This is why we introduce the function $e(\alpha)$ to be determined via the torsion constraint.

Using our formula \eqref{eq:GenFrameAnsatz}, we can specify a generalised frame:
\begin{equation}
E = L \cdot Z \cdot S \cdot \mathring{e}^{-1}\,.
\end{equation}
The $H$-invariant fluxes are parametrised by four functions of $\alpha$ which correspond to the three $H$-invariant fluxes and a factor corresponding to the trombone symmetry. More precisely we have
\begin{equation}
S = \exp\left(d_1(\alpha) t_{[1237]} + d_2(\alpha) t_{[4567]}\right)\cdot \exp\left(c(\alpha) t_{7}{}^8\right)\, \cdot \,\rho(\alpha)\, (\rho(\alpha)\, \mathbb{1}_7)_{\mathcal{E}}\,.
\end{equation}
Where $(\mathbb{1}_7)_{\mathcal{E}}$ is the embedding of the identity matrix in $\GL(7)_\mathcal{E}$.

Finally, elements $Z\in E_{comp}^H$ are all of the form
\begin{equation}
(Z)_M{}^N = \lambda^{-3/4}(\alpha)\, z_M{}^N\,.
\end{equation}
where
\begin{equation}
(z^{-1})_{\Lambda}{}^\Sigma = \lambda(\alpha)^{-3/8}\begin{pmatrix}
	0&\lambda(\alpha)&0&0&0&0&0&0\\
	0&0&\lambda(\alpha)&0&0&0&0&0\\
	0&0&0&\lambda(\alpha)&0&0&0&0\\
	0&0&0&0&0&-1&0&0\\
	0&0&0&0&0&0&-1&0\\
    0&0&0&0&0&0&0&-1\\
	1&0&0&0&\lambda(\alpha)&0&0&0\\
	0&0&0&0&1&0&0&0\\
\end{pmatrix}
\end{equation}
as an $\SL(8)$ matrix in the fundamental representation. You can check that $H_D =( E^\flat)^{-1} H_\mathcal{E} E^\flat$, as expected from our constraints on $E^\flat$.

With this ansatz, depending only on the five functions $\rho,\,d_1,\,d_2,\,c$ and $e$ of the base, we can compute the torsion constraint \eqref{eq:TorsionConstraintsFinal}. The first of such constraints is
\begin{equation}
d(\mathbb{P}_A{}^M E_{M\,[mn]} dx^m \wedge dx^n) = 0\,,
\end{equation}
which fixes that
\begin{equation}
\begin{array}{ll}
	d_1(\alpha) = \frac{C_1 (1+ \lambda^2(\alpha))}{\lambda^{3/2}(\alpha)}\,, 
	&d_2(\alpha) = \frac{C_2 (1+ \lambda^2(\alpha))}{\lambda^{3/2}(\alpha)}\,, \\
	\rho(\alpha) = \frac{g\, \lambda^{1/2}(\alpha)}{(1+\lambda^2(\alpha))^{1/3}}\,,
	&e(\alpha) = \frac{\lambda'(\alpha)}{1+\lambda^2(\alpha)}\,.
\end{array}
\end{equation}
The second differential constraint
\begin{equation}
\partial_m (\mathbb{P}_A{}^M E_{M}^{[mn]}) = 0
\end{equation}
imposes that
\begin{equation}
c = \sqrt{3}\lambda(\alpha) + k\,\frac{(1+\lambda^2(\alpha))^2}{\lambda^3(\alpha)}\,.
\end{equation}
Our final answer then seems to depend on four constants $C_1,\,C_2,\,g$ and $k$ as well as a function of the base $\lambda(\alpha)$. However, using a coordinate redefinition, we can fix $e(\alpha) = 1$, which imposes $\lambda(\alpha ) = \tan(\alpha)$. The rescaling factor $g$ can be fixed to one, it corresponds to fixing the trombone symmetry of M-theory. Finally, the constants $C_1$, $C_2$ and $k$ can be sent to zero by using the gauge transformations of M-theory.

In this simplified form, we observe that the 11D metric at the origin of scalar space is that of the round seven sphere, the 3-form is vanishing and its dual-six form is a primitive of the volume form on $S^7$. This shows that the four-dimensional supergravity we considered can only be uplifted to $S^7$ and thus only appear as a $\mathbb{Z}_2$ consistent truncation of the $\SO(8)$ maximal gauged supergravity. As such, we will not present the full uplift for this very simple consistent truncation as it can always be extended to a consistent truncation to the full $\SO(8)$ gauged maximal supergravity \cite{deWit:1986oxb,Varela:2015ywx}.

\section{Type IIB embeddings of pure \texorpdfstring{$\mathcal{N}=4$}{N=4} supergravity}
\label{sec:TypeIIBExample}

A more interesting example consists in classifying the uplifts of pure $\SO(4)_R$-gauged four-dimensional $\mathcal{N}= 4$ supergravity in type IIB \cite{Schon:2006kz,DallAgata:2023ahj}. The choice of gauging is specified by requiring the existence of an $\mathcal{N}=4$ supersymmetric AdS$_4$ vacuum \cite{Louis:2012ux}.

\paragraph{The pure $\mathcal{N}=4$ gauged supergravity}
The duality group of pure $\cN=4$ supergravity is $G_D = \SO(6)_D \times \SL(2)_D$\footnote{We label these group with the subscript $_D$ in order to differentiate them from other $\SO(6)$ and $\SL(2)$ groups present in our discussion.}. The bosonic sector of this theory contains a metric, $g$, six electric vectors and their six dual magnetic vectors fields, jointly denoted  $A(x)^A$, as well as 2-forms in the adjoint representations of $G_D$. Finally, the theory contains a complex scalar field  $\tau = - \chi + i\, e^\xi$ parametrising the coset space $\SL(2)_D/\SO(2)$. We introduce the following indices: $_I = 1,\,\dots,\,6$ labels the fundamental $\SO(6)_D$-representation while $_\pm$ labels the fundamental $\SL(2)_D$-representation.

Following the discussion in \cite{Louis:2012ux}, in order to admit a maximally supersymmetric AdS$_4$ vacuum, we must select the gauge group $G_g=\SO(4) \cong\SO(3)_1 \times \SO(3)_2$. The group $\SO(3)_1$ is gauged using electric vectors while the group $\SO(3)_2$ is gauged using magnetic ones. To specify the embedding tensor, we split the $\SO(6)_D$ fundamental index $I = (i,\,\underline{i})$ where $i = 1,\,2,\,3$ and $\underline{i} = 4,\,5,\,6$ corresponding to the fundamental $\SO(3)_{1}$ or $\SO(3)_2$ representations respectively. As such the vector index $A = I \pm =(i+,\,i-,\,\underline{i}+,\,\underline{i}-)$. The non-vanishing components of the embedding tensor read
\begin{equation}
\begin{split}
    &X_{i+\,k+}{}^{k+} = \epsilon_{ijk}\,, \hspace{5mm} X_{i+\,j-}{}^{k-} = \epsilon_{ijk}\,,\\
    &X_{\underline{i}-\,\underline{j}-}{}^{\underline{k}-} = \epsilon_{\underline{ijk}}\,, \hspace{5mm} X_{\underline{i}-\,\underline{j}+}{}^{\underline{k}+} = \epsilon_{\underline{ijk}}\,.
\end{split}
\end{equation}
We must specify how the scalar couple to the vectors by embedding the scalar manifold in $\mathrm{Sp}(12)$. This is done by defining
\begin{equation}
    \mathcal{M} = -\begin{pmatrix}
         e^\xi & 0 & -\chi\,e^\xi  & 0 \\
        0  & e^\xi& 0 &-\chi\,e^\xi \\
       -\chi\,e^\xi & 0 & e^\xi|\tau|^2 & 0 \\
        0 &-\chi\,e^\xi & 0 & e^\xi |\tau|^2\\
    \end{pmatrix} \otimes \mathbb{1}_{3\times 3}\,.
\end{equation}
Finally, we must introduce 2-forms transforming in the adjoint of $G_D$. Most of them decouple from the equations of motion and only $B_{ij} \in \mathfrak{so}(3)_1$ and $B_{\underline{{ij}}} \in \mathfrak{so}(3)_2$ remain. From these, we can define gauge invariant field strengths:
\begin{equation}
\begin{split}
    \mathcal{H}^{i+} &= dA^{i+} + \frac{1}{2} \epsilon_{jk}{}^{i}A^{j+}A^{k+}\,,\\
    \mathcal{H}^{\underline{i}+}& = dA^{\underline{i}+} +\frac{1}{2}\epsilon_{\underline{jk}}{}^{\underline{i}} A^{\underline{j}-} A^{\underline{k}+}+\frac{1}{2} B^{\underline{jk}}\epsilon_{\underline{jk}}{}^{\underline{i}}\,,\\
    \mathcal{H}^{i-}& = dA^{i-} +\frac{1}{2}\epsilon_{jk}{}^{i} A^{j+} A^{k-}+ \frac{1}{2} B^{jk}\epsilon_{jk}{}^{i}\,,\\
    \mathcal{H}^{\underline{i}-}&= dA^{\underline{i}-} + \frac{1}{2} \epsilon_{\underline{jk}}{}^{\underline{i}}A^{\underline{j}-}A^{\underline{k}-}\,.
\end{split}
\end{equation}
There is also a field strength for the two form which schematically reads as:
\begin{equation}
\begin{split}
    \mathcal{H}^{IJ} =& DB^{IJ} - (\mathbb{P}_{\mathfrak{so}(6)_D})_{AB}{}^{IJ} A^A (d A^B + \frac{1}{3} X_{CD}{}^B A^C A^D)\,.
\end{split}
\end{equation}
The four dimensional equations of motion are
\begin{equation}
\begin{split}
    &\star \mathcal{H}^A= - (\mathbb{C} \mathcal{M} \mathcal{H})^A\,,\\
    &\mathcal{H}^{IJ}= 0\,,\\
    &\square \xi - e^{2\xi} \partial_\mu \chi \partial^\mu \chi - \partial_\xi V + \mathcal{H} \partial_\xi\mathcal{M} \mathcal{H} = 0\,,\\
    &\nabla_\mu\left(e^{2\xi} \partial^\mu \chi\right) - \partial_\chi V + \mathcal{H} \partial_\chi\mathcal{M} \mathcal{H} =0\,,
\end{split}
\end{equation}
where the scalar potential reads
\begin{equation}
    V(\xi,\,\chi)= \frac{1}{6}\,\text{Tr}(\mathcal{M}) - 4 \,.
\end{equation}
These have to be supplemented by the Einstein equations and the Bianchi identities for the field strengths.

\paragraph{The structure group} Following the discussion in \cite{Cassani:2019vcl}, the structure group corresponding the truncation to pure $\cN = 4$ supergravity is $\SU(4)_S$. Following the group theoretical arguments in \cite{Malek:2017njj}, we show how the fundamental representation of $\Es$ branches under this structure group:
\begin{equation}
\label{eq:fundbranchN4IIB}
\begin{array}{ccccc}
E_{7(7)} &\rightarrow& \SO(6,\,6) \times \SL(2)_D &\rightarrow & \SU(4)_S \times \SU(4)_R \times \SL(2)_D\,,\\[2mm]
 \mathbf{56} &\rightarrow & (\mathbf{12},\,\mathbf{2}) \oplus (\mathbf{32'},\,\mathbf{1}) &\rightarrow&(\mathbf{6},\,\mathbf{1},\,\mathbf{2}) \oplus (\mathbf{1},\,\mathbf{6},\,\mathbf{2})  \oplus (\mathbf{4},\,\mathbf{\bar{4}},\,\mathbf{1}) \oplus (\bar{\mathbf{4}},\,\mathbf{4},\,\mathbf{1}) \,.
\end{array}
\end{equation}
Here, the group $\SU(4)_S \times SU(4)_D$ is understood as the maximally compact subgroup of $\SO(6,\,6)$. We will interchangeably use the groups $\SO(6)_S$ and $\SU(4)_S$ since both share the same Lie algebra. From this analysis, we observe first that this structure group is specified by exactly twelve invariant, nowhere vanishing, linearly independent sections 
\begin{equation}
    K_A := K_{I\pm} \hspace{1cm}\text{ where }\hspace{1cm} I= 1,\,\cdots,\,6 \,.
\end{equation}
They correspond precisely to the twelve vectors of the reduced theory. We denote the projector $\mathbb{P}_{I\pm}{}^M$ of the $\mathbf{56}$ on its $G_S$-invariant subspace and we impose the standard condition \cite{Malek:2017njj}
\begin{equation}
    \mathbb{P}_{I\pm}{}^M \mathbb{P}_{J\pm}{}^N \Omega_{MN} = \delta_{IJ} \epsilon_{\pm\pm} \,.
\end{equation}

\paragraph{The internal manifold} The local structure of the internal manifold is specified by the choice of a principal orbit type, $H$, a subgroup of the gauge group $G_g$. The only principal stabiliser admitting an uplift to type IIB is $H=\SO(2)_1 \times \SO(2)_2$. Indeed, principal stabiliser are in one to one correspondence with solutions to the section constraint and one such solution was built in \cite{Guarino:2024gke}. This choice of isotropy group leads to an internal manifold of the form 
\begin{equation}
    \Mint = S^2 \times S^2 \times \Sigma\,,
\end{equation}
where $\Sigma$ is a surface and, without loss of generality, we can add a complex structure on $\Sigma$. We will use $z$ as a complex coordinate on $\Sigma$. Finally, one can check that  $H^3(S^2\times S^2) = H^5(S^2\times S^2) =H^7(S^2\times S^2) = 0$ which simplifies the torsion constraint to \eqref{eq:RedTorsionConstraints}.

\subsection{The ansatz} 

\paragraph{Compatible solutions to the section constraint}

Following our method, we start by classifying the solutions to the section constraint $\mathcal{E}_M{}^m$ which are compatible, i.e. such that 
\begin{equation}
    \mathbb{P}_A{}^M \mathcal{E}_M{}^{\underline{m}} = \Theta_A{}^{\underline{m}}\,.
    \label{eq:SectionConstrainttheta}
\end{equation}
The reduced embedding tensor $\Theta_A{}^{\underline{m}}$ is
\begin{equation}
    \Theta_{1+}{}^1 = \Theta_{2+}{}^2 = \Theta_{\underline{1}-}{}^{\underline{1}} = \Theta_{\underline{2}-}{}^{\underline{2}} = 1\,,
\end{equation}
and all the other entries vanish. According to the branching \eqref{eq:fundbranchN4IIB}, we have
\begin{equation}
    \mathcal{E}_M = (\mathcal{E}_{I\pm},\,\mathcal{E}_{\underline{I}\pm},\,\mathcal{E}_{{}_\mu{}^{\underline{\mu}}},\,\mathcal{E}_{{}_{\underline{\mu}}{}^{\mu}})\,,
\end{equation}
where $\underline{I}$ labels the fundamental of $\SO(6)_S$ while $\mu$ and $\underline{\mu}$ label the spinorial representations of $\SU(4)_D$ and $\SU(4)_S$ respectively. Obviously we must fix $\mathcal{E}_{I\pm}{}^{\underline{m}} = \Theta_{A}{}^{\underline{m}}$. With this ansatz, one needs to solve the section constraint equations $Y^{MN}{}_{PQ} \mathcal{E}_M{}^{\underline{m}}\mathcal{E}_N{}^{\underline{n}} = 0$. However, by definition of $H_{\mathcal{E}}$, we have the relation 
\begin{equation}
    (h_\mathcal{E})_M{}^N \mathcal{E}_N{}^n h_{n}{}^m = \mathcal{E}_M{}^n\,.
    \label{eq:hInvSectionConstraints}
\end{equation}
This can be used to simplify our ansatz for $\mathcal{E}$ by specifying an embedding of the isotropy group $H$ in $\Es$, i.e. a map $H\rightarrow \SU(4)_S$. Since, for a given type of solution to the section constraint this map is unique when it exists, and we have an example of such a map from \cite{Guarino:2024gke}, we must embed $H$ diagonally in $\SU(4)_S \times \SO(6)_D$. Then, we classify H-singlets in the $\mathbf{56} \times (\mathbf{2} + \mathbf{2} + 2\cdot \mathbf{1}) $. There are $32 + 2\times 16$ such singlets, all other contributions must be set to zero. 

Solving partially the section constraint, we can use an $\SU(4)_S$ gauge fixing to impose that 
\begin{equation}
    \mathcal{E}_{I\pm} = \mathcal{E}_{\underline{I}\pm} = \Theta_{I\pm}\,.
\end{equation}
And we only have to solve for the fermionic $\SU(4)_R$ representations: $\mathcal{E}_{{}_\mu{}^{\underline{\mu}}}$ and $\mathcal{E}_{{}_{\underline{\mu}}{}^{\mu}}$. 

Although the exact solution is not particularly enlightening, we present some properties of it here. In particular the space of $H$-invariant compatible solutions to the section constraint can be described as
\begin{equation}
    \mathcal{E}^H_{\text{comp}} \cong \SO(2,\,2)/\SO(1,\,2) \times \GL(2)_\Sigma\,.
\end{equation}
The $\GL(2)_\Sigma$ factor simply correspond to a choice of vielbein on the basis $\Sigma$. However, the $\SO(2,\,2)/\SO(1,\,2)$ factor induces a genuine action on $\mathcal{E}^{\underline{m}}$. This group can be build by considering the $H$-invariant subgroup of $\SO(6,6)$:
\begin{equation}
    \SO(6,\,6) \rightarrow (\SL(2)_1 \times \SO(2)_1) \times (\SL(2)_2 \times \SO(2)_2) \times \SO(2,\,2)\,.
\end{equation}
The subgroup leaving $(\mathcal{E}_{I\pm},\,\mathcal{E}_{\underline{I}\pm})$ invariant is $\SO(1,\,1)_1 \times \SO(1,\,1)_2 \times \SO(2,\,2)$. The first two $\SO(1,\,1)$ factors leave the full section constraint invariant and are thus part of $\mathcal{S}_\mathcal{E}$ for any compatible section constraint. However, the $\SO(2,\,2)$ factor turns out to be the stabiliser of the $H$-invariant compatible solution to the section constraint! Finally, since $\SO(2,\,2)/\SO(1,\,2) \cong \SL(2)$, we have an $\SL(2) \times \GL(2)_\Sigma$ family of solutions to the section constraint.

\paragraph{The frame ansatz}

Using the results presented above, the most generic frame corresponding to a reduction of the structure group with the appropriate torsion must be of the form
\begin{equation}
    E = L\cdot \, E^\flat\, \cdot\, e^{-1}\,.
\end{equation}
Here, $L$ is the $S^2 \times S^2$ coset representative while the vielbein $e = \begin{pmatrix} \mathring{e} &0 \\0 & e_{\Sigma}\end{pmatrix}$ built from the round $S^2\times S^2$ vielbein and a generic vielbein on $\Sigma$. The flat frame $E^\flat$, which is a function of $\Sigma$, is the product of an element of $\SL(2)$ with the $H$-invariant fluxes in $\mathcal{S}_{\mathcal{E}}$. These fluxes contain the trombone scaling, the type IIB scalars, three pairs of $2$-form fluxes, three $4$-form fluxes as well as two dual six-form fluxes. One can simplify this ansatz by using coordinate transformations on $\Sigma$ to fix $e_\Sigma$ to be conformal. Moreover, the torsion does not depend on two of the 2-forms and one of the 4-form fluxes which can be set to zero using gauge transformations.

\paragraph{The differential conditions}

The last step consists in solving eq. \eqref{eq:TorsionConstraintsFinal} which is a series of PDE on $\Sigma$\footnote{It turns out that it was computationnaly simpler to classify all (co-)closed equivariant forms on $\Mint$ to provide an ansatz for $K_A$ and then compare this to $(\mathbb{P} E)_A$. The two methods are equivalent.}. Obviously, the vector part of the non-vanishing sections $K_A$, $K_A{}^{m} = k_A{}^m$, are the standard $\mathfrak{so}(3) \oplus \mathfrak{so}(3)$ Killing vectors on the two-spheres. Introducing the usual spherical coordinates on each $S^2_{i}$, $(\theta_i,\,\phi_i)$, we write
\begin{equation}
\begin{array}{ll}
    k_1 = \partial_{\phi_1}\,, & k_{\underline{1}} = \partial_{\phi_2}\,,\\
    k_2 = \cos\phi_1\, \partial_{\theta_1} -\cot\theta_1\,\sin\phi_1 \partial_{\phi_1}\,,& k_{\underline{2}} =  \cos\phi_2\, \partial_{\theta_2} -\cot\theta_2\,\sin\phi_2 \partial_{\phi_2}\,, \\
    k_3 = -\sin\phi_1  \partial_{\theta_1} - \cot\theta_1 \cos \phi_1 \partial_{\theta_1}\,,\hspace{1cm} & k_{\underline{3}}=  -\sin\phi_2  \partial_{\theta_2} - \cot\theta_2 \cos \phi_2 \partial_{\theta_2}\,,
\end{array}
\end{equation}
and we must impose that $k_{i+} = k_i$ and $k_{\underline{i}-} = k_{\underline{i}}$. We also introduce the embedding coordinates of the two-spheres in $\mathbb{R}^3\times \underline{\mathbb{R}}^3$ as
\begin{equation}
    \begin{array}{ll}
    Y^1 = \cos(\theta_1)\,,& Y^{\underline{1}} = \cos(\theta_2)\,,\\
    Y^2 =\sin(\theta_1)\, \sin(\phi_1) \,,& Y^{\underline{2}} = \sin(\theta_2)\,\sin(\phi_2)\,,\\
    Y^3 =\sin(\theta_1)\,\cos(\phi_1) \,,& Y^{\underline{3}} =\sin(\theta_2) \,\cos(\phi_2)\,.
    \end{array}
\end{equation}
They satisfy the relations $k_i (Y^j) = \epsilon_{ijk} Y^k$. Following the notations of \cite{Assel:2011xz}, the final solution to both torsion and compatibility constraints should depend on two harmonic functions $h_1$ and $h_2$ as well as their harmonic conjugate $h_1^D$ and $h_2^D$. We write 
\begin{equation}
    2 \mathcal{A}_1 = h_1^D + i\,h_1\hspace{5mm}\text{ and }\hspace{5mm} 2 \mathcal{A}_2 = h_2 - i\,h_2^D\,,
\end{equation}which are both holomorphic functions on $\Sigma$. Equivalently, we have that
\begin{equation}
    \begin{array}{ccc}
    h_1 = -i(\mathcal{A}_1 - \bar{\mathcal{A}}_1) & & h_1^D = \mathcal{A}_1 + \bar{\mathcal{A}}_1,\,\\
    h_2= \mathcal{A}_2 + \bar{\mathcal{A}}_2 && h_2^D = i(\mathcal{A}_2 -\bar{\mathcal{A}}_2)\,.
    \end{array}
\end{equation}
This is exactly the result we obtain by solving the torsion constraint. The solution, expressed in terms of the invariant sections $K_A$ reads
\begin{equation}
\label{eq:sectionFinalIIB}
\begin{array}{clll}
    &K_{i+} = &k_{i}{} &+ d(h_2^D Y^i)^+\\
    &&&- 
 d\left(\,(2h_1 h_2 + 2h_1^D h_2^D - \lambda) \,\text{vol}_2\,Y^i \right) \\
 &&&-  \smallstar d\left(\epsilon_{ijk} Y^j dY^k \wedge \text{vol}_2 \wedge\left(( 2 (h_1^D h_2^D - h_1 h_2)+ \lambda)  dh_2^D + 4 h_2 h_2^D dh_1 \right)\right)^-\\ &&&+0\,,\\[2mm]
    &K_{\underline{i}+} = &0 &+ d(-h_2 Y^{\underline{i}})^+\\
    &&&+ d\left(-4\epsilon_{\underline{ijk}} Y^{\underline{j}} dY^{\underline{k}} \wedge (h_2 dh_1^D- h_1^D dh_2) \right) \\
    &&&-  \smallstar d\left(\epsilon_{\underline{ijk}} Y^{\underline{j}} dY^{\underline{k}} \wedge \text{vol}_1 \wedge\left((2 ( h_1 h_2 -h_1^D h_2^D)-\lambda)  dh_2 + 4 h_2 h_2^D dh_1^D \right)\right)^-\\
    &&& + 8 \sqrt{g_{S^2\times S^2}} \left(2 h_1 h_2 + 2 h_1^D h_2^D - \lambda\right) (\partial h_1 \bar\partial h_2 + \bar\partial h_1 \partial h_2) k^*_i\,,\\[3mm]
   &K_{i-} = &0 &+ d(-h_1 Y^i)^- \\
   &&&+d\left(4\,\epsilon_{ijk} Y^{j} dY^{k} \wedge (h_1 dh_2^D - h_2^D dh_1)\right)\\
   &&&- \smallstar d\left(\epsilon_{ijk} Y^j dY^k \wedge \text{vol}_2 \wedge\left(( 2 (-h_1^D h_2^D + h_1 h_2)+ \lambda)  dh_1 + 4 h_1h_1^D dh_2^D \right)\right)^+\\
    &&& - 8 \sqrt{g_{S^2\times S^2}} \left(2 h_1 h_2 + 2 h_1^D h_2^D + \lambda\right) (\partial h_1 \bar\partial h_2 + \bar\partial h_1 \partial h_2) k^*_{\underline{i}}\,,
   \\[2mm]
   & K_{\underline{i}-} = &k_{\underline{i}}{} &+ d(-h_1^{D} Y^{\underline{i}})^- \\
   &&&+ d\left( \,(2 h_1 h_2 + 2 h_1^D h_2^D + \lambda)\,\text{vol}_1\,Y^{\underline{i}} \right) \\
   &&&-  \smallstar d\left(\epsilon_{\underline{ijk}} Y^{\underline{j}} dY^{\underline{k}} \wedge \text{vol}_1 \wedge\left((2 ( h_1 h_2 -h_1^D h_2^D)+\lambda)  dh_1^D + 4 h_1 h_1^D dh_2 \right)\right)^+\\ &&&+0\,,
\end{array}
\end{equation}
where $\lambda$ is a real harmonic function\footnote{It can also be characterised as the real part of the holomorphic function $l$ satisfying $\tfrac{i}{2} \partial l = \mathcal{A}_1 \partial \mathcal{A}_2 -  \mathcal{A}_2 \partial  \mathcal{A}_1$. Requiring $\lambda$ to be real just fixes the gauge $\lambda(z,\,\bar{z}) \rightarrow \lambda(z,\,\bar{z}) + f(\bar{z})$.} such that
\begin{equation}
    \frac{i}{4}\,\partial \lambda =  \mathcal{A}_1 \partial  \mathcal{A}_2 -  \mathcal{A}_2 \partial  \mathcal{A}_1\,.
\end{equation}
 The 1-forms $k^*_i$ and $k^*_{\underline{i}}$ are the dual to the Killing vectors on $S^2 \times S^2$ with respect to the $\SO(3)_1\times\SO(3)_2$-invariant metric given by the vielbein $\mathring{e}$. 

We have explicitly verified that both the compatibility constraints and the torsion constraint are satisfied for the sections \eqref{eq:sectionFinalIIB}. We also checked that the frame is globally well defined by computing its determinant to obtain 
\begin{equation}
    \text{det}(E)^{1/28} = \text{cste} \,\times\,\,h_1 h_2 \,\partial\bar{\partial}(h_1 h_2)\,.
\end{equation}
Following \cite{DHoker:2007hhe}, the quantity $h_1 h_2 \partial \bar\partial(h_1 h_2)$ is always strictly negative and vanishes only for specific points corresponding to back-reacted brane singularities at the boundary of the base. As such, we have provided a consistent truncation arbitrarily close to brane sources.

\subsection{The consistent truncation in supergravity}

Plugging our solution in the frame ansatz, we immediately obtain the generalised metric of ExFT. Using the ExFT dictionnary, this yields the various fields of type IIB supergravity. The ExFT dictionary requires to perform a series of non-linear field redefinitions to obtain the IIB fields whose equations of motion reduce to those of 4 dimensional $\mathcal{N} = 4$ supergravity. This dictionary is worked out in appendix \ref{app:KKCalculus}. Here we write down the consistent truncation in a supergravity language, for any sections parametrised by $h_1$ and $h_2$.

\subsubsection{The dictionary}

First, we introduce the KK vectors $A_{KK} = A^A(x^\mu)\, \iota_{(\Theta_A{}^\alpha k_\alpha)}$ which will twist the metric and the various $p$-forms. This is done by introducing the operator
\begin{equation}
    e^{A_{KK}} = \sum_{n=0}^{\infty} \frac{(A^A \iota_{(\Theta_A{}^\alpha k_\alpha)})^n}{n!}\,.
\end{equation}
This operators allows us to build the type IIB supergravity fields $C_{(p)}$ from the ``KK-flat" fields $\bar{C}_{(p)}$. The definition is
\begin{equation}
    C_{(p)} = e^{A_{KK}} \bar{C}_{(p)}
\end{equation}
and the ExFT dictionary gives a prescription for the $\bar{C}_{(p)}$ fields. In coordinates, acting with $e^{A_{KK}}$ is equivalent to the more familiar replacement
\begin{equation}
    dy^m \rightarrow Dy^m = dy^m + A^A \Theta_A{}^\alpha k_\alpha{}^{m}\,,
\end{equation}
for any of the internal coordinates $y^m$.

Then, following the standard definitions from ExFT, $p$-forms and the metric are split into various contributions, organised by their rank in the external space. For example
\begin{equation}
    \bar{C}_{(p)} = c + A^A c_A + B^{AB} c_{AB} + \cdots
\end{equation}
where $c$ is the scalar contribution, read from the generalised metric $\mathcal{M}_{MN}(x,\,Y^M)$ (for the precise expressions in the type IIB context, see \cite{Inverso:2016eet}). The vector contributions $A^A c_A$ are given by reading the appropriate components of the sections $K_{A\,(p-1)}$. The ``$\cdots$" represent further non-linear field redefinitions needed to preserve gauge invariance. Details are available in the appendix \ref{app:KKCalculus}.

\subsubsection{Final formulas}
To present the final formulas in a compact way, we start by defining a symmetric ``product" and an anti-symmetric one:
\begin{equation}
\begin{array}{rl}
    \scalP{a}{b} =& \partial a\bar{\partial}b + \bar\partial a{\partial{b}} = 2 \text{Re}\left( \partial a\bar{\partial}b\right) = 2\text{Re}\left( \bar\partial a{\partial}b\right)\,,\\
    a\wedge b =& -i(\partial a \bar\partial b-\partial b \bar\partial a )= 2 \text{Im}\left(\partial a\bar{\partial}b\right)=- 2 \text{Im}\left(\bar\partial a {\partial}b\right)\,.
\end{array}
\end{equation}
We define
\begin{equation}
    \begin{array}{rl}
    w =& \scalP{\log h_1}{\log h_2}\,, \\
    n_1 =&  \frac{\scalP{\log{h_1}}{\log{h_1}}}{\scalP{\log h_1}{\log h_2}}- e^\xi |\tau|^2\,,\\
    n_2 =&  \frac{\scalP{\log{h_2}}{\log{h_2}}}{\scalP{\log h_1}{\log h_2}}- e^\xi\,,\\
    n_0 =& \frac{\scalP{\log{h_1}}{\log{h_1}}}{\scalP{\log h_1}{\log h_2}}-e^{-\xi}\,.
    \end{array}
\end{equation}
The quantities $(n_1,\,n_2,\,w)$ are functions of $h_{1,\,2}$ invariant under constant rescalings of $h_{1,\,2}$. The two functions $n_i$ are scalars while $w$ transforms as a $(1,1)$-form on $\Sigma$. Upon fixing $\tau = i$, these functions are rescalings of the functions ($N_1, N_2,\,W$) of \cite{Assel:2011xz}.

Using these notations, we have the type IIB axio-dilaton fields:
\begin{equation}
\begin{array}{rl}
    \text{exp}(\Phi) =& e^\xi\,\frac{h_1 }{h_2}\frac{n_0}{\sqrt{n_1  n_2}}\,,\\[2mm]
    C_0 =& \chi\frac{\,h_1\wedge h_2}{h_1^2\, n_0\, w}\,.
\end{array}
\end{equation}
The KK-flat metric reads
\begin{equation}
    \bar{ds}^2 = \Delta^{-1} \left(ds^2_{\text{ext}} - n_1^{-1} ds^2_{S^2_1}- n_2^{-1} ds^2_{S^2_2} - 2 w \,dz \,d\bar{z} \right)
\end{equation}
and the warping factor is
\begin{equation}
    \Delta^{-4} = 16 n_1 n_2 h_1^2 h_2^2\,.
\end{equation}
The two 2-forms read
\begin{equation}
    \begin{array}{rl}
    \bar{B}_2 =& -2\chi \,e^\xi\frac{h_1}{n_1}\text{vol}_{S_1^2} + 2 h_1^D \text{vol}_{S_2^2} +2 h_1\frac{(\log h_1 \wedge \log h_2) \,}{n_2 \,w}\,\text{vol}_{S_2^2} \\[3mm]
    &+ A^{i-} d( h_1 \,Y^i)+ A^{\underline{i}-}d(h_1^D \,Y^{\underline{i}})\\[3mm]
    &+(B^{ij} +\frac{1}{2} A^{i+} A^{i-})\,\epsilon_{ijk}\, h_1 Y^k +\frac{1}{2} A^{\underline{i}-} A^{\underline{j}-} \epsilon_{\underline{ijk}} \, h_1^D Y^{\underline{k}}\,, \\[3mm]
    \bar{C}_2 =& 2 h_2 ^D \text{vol}_{S_1^2} - 2h_2\frac{(\log h_1\wedge \log h_2)}{n_1 \,w}\,\text{vol}_{S_1^2} + 2 \chi e^{\xi}\frac{h_2}{n_2} \text{vol}_{S_2^2}\\[3mm]
    &+ A^{i+} d(h_2^D \,Y^i)+ A^{\underline{i}+}d(h_2 \,Y^{\underline{i}})\\[3mm]
    &+ A^{i+} A^{j+} \epsilon_{ijk}\, h_2^D Y^{k}+(B^{\underline{ij}} +\frac{1}{2} A^{\underline{i}-} A^{\underline{i}+})\,\epsilon_{\underline{ijk}}\, h_2 Y^{\underline{k}}\,.  \\[3mm]
\end{array}
\end{equation}
Finally, the improved 5-form is
\begin{equation}
\label{eq:F5final}
\begin{split}
  \bar{\tilde{F}}_5   =&\phantom{+}\frac{1}{n_1\,n_2}\text{vol}_{S^2_1 \times S^2_2} \wedge\left(\star_\Sigma dj + 4\,\left(\,(1 - e^\xi)\, h_2 dh_1 - (1-e^\xi  |\tau|^2)\, h_1 dh_2\right)\right)\\
  &  +  \text{vol}_{\text{ext}} \wedge  \left(dj + 4\,\left(\,(1 - e^\xi)\, h_2 dh_1^D - (1-e^\xi |\tau|^2)\, h_1 dh_2^D\right)\right)\\
    &+ \,4\, \frac{h_1 h_2}{n_1 n_2} ( d\xi- e^{2\xi}  \chi d\chi)\wedge\text{vol}_1 \wedge \text{vol}_2\\
    &+ \,8 w \,h_1 h_2 \star_{\text{ext}}(d\xi- e^{2\xi}  \chi d\chi)\wedge\text{vol}_{\Sigma}\\
    &+4(1+\bar\star)\Big{[}\mathcal{H}^{i+} (-Y^i h_1 dh_2 \wedge \text{vol}_2 n_2^{-1})\\
    &\phantom{+4(1+\star)\Big{[}}+\mathcal{H}^{\underline{i}+}(- Y^{\underline{i}} h_1 dh_2^D \wedge \text{vol}_2 n_2^{-1}- i\, w h_1 h_2\, \epsilon_{\underline{ijk}}Y^{\underline{j}} dY^{\underline{k}} \wedge dz\wedge d\bar{z}) )\\
    &\phantom{+4(1+\star)\Big{[}} + \mathcal{H}^{i-}(- Y^{i} h_2 dh_1^D \wedge \text{vol}_1 n_1^{-1} - i\, w h_1 h_2\epsilon_{ijk}Y^{j} dY^{k}\wedge dz\wedge d\bar{z}) )\\
    &\phantom{+4(1+\star)\Big{[}}+\mathcal{H}^{\underline{i}-} (Y^{\underline{i}}h_2 dh_1\wedge \text{vol}_1 n_1^{-1})
    \Big]\\
    &+\mathcal{H}^{MN}X_{MN}{}^P(\cdots)\,,\\
\end{split}
\end{equation}
where
\begin{equation}
    j = - 3 \lambda^D - 12 (\bar{ \mathcal{A}_1}  \mathcal{A}_2 +  \mathcal{A}_1 \bar{ \mathcal{A}_2}) - 4 \frac{h_1 \wedge h_2}{w}
\end{equation}
and $\lambda^D$ is the harmonic conjugate to $\lambda$. The 4-form $\text{vol}_{\text{ext}}$ is the volume form of the metric $ds^2_{\text{ext}}$. We have not written down the specific terms coupling to $\mathcal{H}^{MN}$ because the self-duality equations of motion ensure that $\mathcal{H}^{MN}X_{MN}{}^P = 0$.

We conclude this section with a few comments. We have added corrections to $\tilde{F}_5$ to ensure its self-duality whenever $\mathcal{H}^M$ satisfies the four-dimensional twisted self-duality equations. We have used the same twisted self-duality equations to remove explicit four-dimensional scalar dependences in the last fives lines of \eqref{eq:F5final}. As a sanity check, we have verified that the Bianchi identity for $\tilde{F}_5$ and the e.o.m. of the type IIB axio-dilaton do reduce to the four-dimensional equations of motion. Finally, for the particular choices, $z= \eta + i \alpha$, $\mathcal{A}_1 = -\tfrac{i}{4\sqrt{2}} \exp(-z)$ and $\mathcal{A}_2 =\tfrac{i}{4\sqrt{2}} \exp(z)$, one would recover the truncation around the S-fold described in \cite{Guarino:2024gke}.

\section{Outlook}

In this paper, we have spelled out how to perform the uplift, when it exists, of any given gauged supergravity. In particular, we have provided new consistent truncations to pure $\mathcal{N}=4$ $D=4$ $\SO(4)$-gauged supergravity around any of the $\mathcal{N}=4$ AdS$_4$ solutions of type IIB. This opens the possibility to study the uplifts of black holes, introduce probe branes, etc., in order to perform new checks of the holographic dictionary. In particular, our construction provides the type IIB realisation of the universal result of \cite{Azzurli:2017kxo}, which relates the free energy of SCFTs to their twisted index. The uplift of the magnetically charged black holes of \cite{Romans:1991nq,Caldarelli:1998hg} in type IIB will explicitly realise the holographic dual of the computations of \cite{Coccia:2020wtk}, in which these universal relations are checked for the $T^\sigma_\rho[SU(n)]$ SCFTs dual to the AdS$_4$ solution we uplifted \cite{Assel:2011xz}.

For the particular type IIB uplift we considered, the principal orbit type is  $\text{U}(1) \times \text{U}(1)$. However, other orbit types are possible. In particular, choosing $\Hprin = \{e\}$, one recovers at least one compatible section from the M-theory uplift presented in section \ref{sec:nv6Example}. This implies that all uplifts with $\Hprin=\{e\}$ are M-theory uplifts. It would be interesting to work out whether or not all M-theory uplifts arise as simple truncations of the $\SO(8)$-gauged supergravity. We have not yet been able to classify these uplifts due to the complexity of the algebraic equations obtained from studying the space of compatible solutions to the sections constraints.

Our formalism opens several research directions. First, one could consider the uplifts of other non-maximal gauged supergravities. For examples, there exists other gaugings of the pure $\mathcal{N}=4$ gauged supergravity, such as the purely electric gauging of \cite{Bergshoeff:1985ms} reformulated in the embedding tensor formalism in \cite{Schon:2006kz}. One could also consider the uplifts of various $\mathcal{N}=3$ gauged supergravities, amongst other the ones considered in \cite{Fre:2022ncd}. Reducing the number of supersymmetries, classifying uplifts of all $\mathcal{N}=2$ supergravities is probably hopeless (given that the internal manifold only admits a $U(1)$ symmetry). However, partial classification for gauged $\mathcal{N}=2$ supergravities with extra vector multiplets, and larger gauge groups, should be possible with our method. It would also be interesting to study whether further reductions of known $G_S$-structures (e.g. to a smaller $H_S$-structure see app. \ref{app:EflatConstraints}) exists, and, when they do, if they admit a constant intrinsic torsion. 

We would also like to study uplifts to massive type IIA supergravity. In mIIA, the definition of the generalised Lie derivative needs to be modified to include a ``flux" corresponding to the Romans mass. This extra term in the generalised Lie derivative modifies our discussion and does not allow one to conclude that the internal manifold is of the form $G_g/H\times B$. Including this deformation, it would be interesting to study uplifts to massive type IIA of the pure $\mathcal{N}=4$ $D=4$ gauged supergravity considered in this paper. This uplift should have a principal orbit type different from the type IIB one ($H= U(1)^2$) or the M-theory one ($H=\{e\}$). In the same vein, we wonder if our methods could be applied to $D<4$ using $\mathrm{E}_{8(8)}$-ExFT or the ExFT based on the infinite dimensional Lie algebras $\mathrm{E}_{9}$ \cite{Bossard:2018utw,Bossard:2021jix}. We leave these and other questions for future work.

\section*{Acknowledgements}
We thank Gianluca Inverso for his valuable contributions during the first stages of this project, and Davide Cassani for insightful discussions. C.S. is supported by a Postdoctoral Research Fellowship granted by the F.R.S.-FNRS (Belgium).

\newpage

\appendix

\section{Index conventions}
\label{app:Conventions}
Due to the amount of indices related to different objects, we compile here a list of the different types of indices we use.
\paragraph{Section \ref{sec:Intro}, \ref{sec:Review} and \ref{sec:Construction}:} In these sections, which are devoted to the systematic uplift of supergravities, the following indices are used:
\begin{itemize}
    \item The indices $M,\,N,\,P,\,\dots$ label the fundamental representations of $\Ed$ groups.
    \item The indices $A,\,B,\,C,\,\dots$ label the $G_S$-singlets. 
    \item The indices $m,\,n,\,p,\,\dots$ label the coordinates on $\Mint$, $y^m$.
    \item The index $m$ splits into tilded indices $\tilde{m}$ and barred indices $\bar{m}$, labelling coordinates on the base $B$ or on the fibre $G_g/H$.
    \item The indices $\underline{m},\,\underline{n},\,\underline{p},\,$ label flattened indices on $\Mint$; this index splits into $\tilde{\underline{m}}$ and $\bar{\underline{m}}$. Incidently, $\bar{\underline{m}}$ can also label elements of a basis of $\mathfrak{K} = \mathfrak{g}_g /\mathfrak{h}$.
\end{itemize}

\paragraph{Section \ref{sec:nv6Example}: M-theory example}

The index of the fundamental $\Es$ representation $M$ splits into $\SL(8)$ representations as $\mathbf{56} \rightarrow \mathbf{28} \oplus \mathbf{28}'$: $M\rightarrow _{[\Lambda \Sigma]} \oplus ^{[\Lambda \Sigma]}$. The indices $\Lambda,\,\Sigma,\,\dots = 1,\,\dots,\,8$ label the fundamental representation of $\SL(8)$.

\paragraph{Section \ref{sec:TypeIIBExample}: Type IIB example}
For the type IIB example, the structure group $G_S = \SU(4)_S$. As before, the $G_S$-singlets are labelled by the index $A,\,B,\,C,\,...$, while the fundamental indices of $\Es$ are labelled by $M,\,N,\,P,\,...$ These indices split under the $\SO(6)_S\times \SO(6)_D\times \SL(2)_D$ group as
\begin{itemize}
    \item $A \rightarrow I \pm$, with $I=1,\,\dots,\,6$ and $\pm$ are the fundamental $\SO(6)_D \times \SL(2)_R$ indices.
    \item $M \rightarrow I\pm \oplus \underline{I}\pm \oplus \mu\bar{\mu} \oplus \bar{\mu} \mu$, where $\underline{I}$ labels the antisymmetric of $\SU(4)_S$ (i.e. the fundamental of $\SO(6)_S$), and $\mu$ and $\bar{\mu}$ label the fundamental of $\SU(4)_D$ and $\SU(4)_S$ respectively (i.e. the spinorial representations of $Spin(6)_D$ and $Spin(6)_S$).
\end{itemize}
These further branch as irreps of the gauge group $G_g = \SO(4)_R \subset \SO(6)_D$:
\begin{itemize}
    \item $I \rightarrow (i,\,\underline{i})$, where $i=1,\,2,\,3$ and $\underline{i}=\underline{1},\,\underline{2},\underline{3}$.
\end{itemize}
When it comes to internal coordinates, $m$ splits into a set of four angular coordinates $(\theta_1,\,\phi_1,\,\theta_2,\,\phi_2)$ on the two two-spheres and a complex coordinate on the base.

\section{Group actions on differentiable manifolds}
\label{app:GroupActions}

In the main text, we use several properties of group actions on differentiable manifolds. Following \cite{Meinrenken2003}, we summarise here these properties and work out a couple of examples relevant for our work. In particular, we will show that for a proper group action $G$ on $M$, the quotient space $M/G$, although singular, can be stratified $M/G = \cup_{(H)} X_{(H)}$. These ``strata" $M_{(H)}$ are labelled by conjugacy classes of subgroups of $G$ and are themselves smooth manifolds. The study of this structure, relevant for the model \eqref{eq:MintIntro}, is the goal of this appendix.

\subsection{Definitions}

\begin{definition}
    Let $M$ be a differentiable manifold and $G$ a Lie group, a left action of $G$ on $M$ is a group homomorphism $L:G\rightarrow \text{Diff}(M): g \mapsto L_g$
    such that 
    \begin{equation}
        G\times M \rightarrow M : (g,\,m) \mapsto L_g(m)
    \end{equation}
    is smooth.
\end{definition}

\begin{definition}
    Let $M$ be a differentiable manifold, and $\mathfrak{g}$ a finite-dimensional real Lie algebra, an action of $\mathfrak{g}$ on $M$ is a Lie algebra homomorphism $\bullet_M:
        \mathfrak{g} \rightarrow \mathfrak{X}(M): \xi \mapsto \xi_M$
such that \begin{equation}\mathfrak{g}\times M \rightarrow TM : (\xi,\,m)\mapsto \xi_M(m)\end{equation} is smooth.
\end{definition}
\noindent It is a well-known result that the action of Lie groups induces an action of the corresponding Lie algebra. The converse is not always true and requires the $\xi_M$ to be complete.
\begin{theorem}
    Let $\xi \rightarrow \xi_M$ be a Lie algebra action on a manifold $M$, this action integrate to the action of the associated simply connected Lie group $G$ if and only if each $\xi_M$ is complete.
\end{theorem}
\noindent Group action on manifolds can be highly pathological and may prevent us from giving much structure to the underlying manifold. To avoid this, we will always require here that group actions are \emph{proper}.
\begin{definition}
    A $G$-action on a manifold $M$ is proper if the map 
    \begin{equation}
        G\times M \rightarrow M\times M : (g,\,m) \mapsto (m,\,g\cdot m)
        \label{eq:defProper}
    \end{equation}
is proper, i.e. the pre-image of compact sets are compact.
\end{definition}
\noindent This requirement ensures that the quotient space $M/G$ with the quotient topology is separated (``Hausdorff"), i.e. one can build two non-intersecting neighbourhoods for any two distinct points.
\begin{example} We recall some basic examples of proper and improper actions:
\begin{itemize}
    \item The action of $\mathbb{R}$ on $T^2$ with an ``\emph{irrational}" slope is not proper.
    \item The action of a compact group $G$ is always proper.
    \item The left action of a Lie group on itself is proper.
    \item The defining action of $\SO(1,\,1)$ on $\mathbb{R}^{1,\,1}$ is \emph{not} proper. The quotient space contains four half-lines connecting at the orbit of the origin. However, it also contains four points corresponding to the light-rays. The quotient topology is not Hausdorff (similar to that of the real line with two origin).
\end{itemize}
\end{example} 

\begin{definition} Let $M$ be a manifold with a $G$-action:
    \begin{itemize}
        \item $G_m = \{g\in G\,|\, g\cdot m = m\}$ is the \emph{stabliser of} $m$,
        \item $\mathcal{O}_m = \{n\in M\, |\, n = g\cdot m\, \text{ for }\,g\in G\}= G\cdot m$ is the ($G$-)orbit of $m$.
    \end{itemize}
\end{definition}
\noindent One can readily observe from these definitions that the stabiliser $G_m$ for a proper group action is always compact (it is the preimage of $(m,\,m)$ under the map \eqref{eq:defProper}). Moreover, if $n\in \mathcal{O}_m$ then $\exists \,g\in G$ s.t. $n = g\cdot m$ and $G_n = g^{-1} \cdot G_m \cdot g$. In other words, any two points in the same $G$-orbit have identical stabiliser up to conjugation. 
\begin{definition}
    For a subgroup $H \subset G$, we denote by $(H)$ the conjugacy class of $H$ in G. We note
    \begin{itemize}
        \item $M^H = \{m\in M \,|\, H \subset G_m\}$, the fixed-points set of $H$,
        \item $M_H = \{m\in M\,|\, H = G_m\}$, the points of isotropy $H$,
        \item $M_{(H)} = \{m\in M\,|\, (G_m)=(H) \}$, the points of orbit type $(H)$.
    \end{itemize}
\end{definition}

\subsection{Slice theorem}
For proper action, one can locally characterise the structure of the manifold via the following result:
\begin{theorem}[Slice theorem for proper action]
    Let $G$ be a proper action on $M$. Let $m\in M$ with stabiliser $H$, and denote $V_m = T_mM/T_m(G\cdot m)$ the \emph{slice representation}. Then, there exists a G-equivariant map $\phi_m: G\times_{H} V_m \rightarrow M$ s.t. $\phi_m([(e,\,0)]) = m$. This map is a diffeomorphism on its image.
\end{theorem}
\begin{remarks}We want to stress the following three points:
\begin{itemize}
    \item The map $\phi_m$ depends on the point $m$. The same is true for the vector space $V_m$ and the stabiliser $H$, which might vary depending on $m$.
    \item The slice representation $V = T_mM/T_m(G\cdot m)$ admits a linear $H$-action. Indeed, $T_mM$ admits a linear $H$-action given by the differential $(dL_h)_m:T_mM \rightarrow T_mM$ since $H$ fixes $m$. Moreover, $T_m(G\cdot m)$ is a $H$-invariant subspace of $T_mM$ which allows one to take the quotient while preserving the $H$-action. 
    \item The map $\phi_m$ does \emph{not} implies that $M \cong G\times_H V$ because $\phi^{-1}_m$ is only defined on $\text{Im}(\phi_m)$ which is, in general, a subset of $M$.
\end{itemize}
\end{remarks}

\begin{example}
    Let $G = \SO(2)$ and $M = S^2$ where $G$ acts on $M$ as the standard rotation around the north-south axis. This action has two fixed points at each pole $p_N$ and $p_S$ of the sphere. Since $\SO(2)$ is compact, the action is proper. There are three inequivalent slices:
    \begin{itemize}
    \item Near a generic point $p \in S^2 \backslash\{p_N,\,p_S\}$. The stabiliser of $p$ is trivial and the slice representation is $ V = T_p(S^2)/T_p(G\cdot p) = T_p(S^2)/T_p(S^1_p) \cong \mathbb{R}$. This implies that there exists a neighbourhood $U_p$ of $p$ such that
\begin{equation}
    U_p \cong \SO(2) \times_e V \cong \SO(2) \times \mathbb{R}\,.
\end{equation}
    The stabiliser of $p$ is trivial and has a trivial action on $\mathbb{R}$.
    \item Near the north pole $p_N$. The stabiliser of $p_N$ is $\SO(2)$, and $T_p(G\cdot p_N) = \{0\}$. Thus $\exists\, U_{N}$ a neighbourhood of $p_N$ diffeomorphic to
    \begin{equation}
        \SO(2)\times_{\SO(2)} T_{p_N}(S^2)/ \langle 0\rangle \cong T_{p_N}(S^2)\,,
    \end{equation}
    which indeed admits an $\SO(2)$-action.
    \item Near the south pole, for which the details are similar to the slice around $p_N$.
    \end{itemize}
We see that the space of orbits $S^2/\SO(2) \cong I$ reflect that structure. The interior of the closed interval $I$ corresponds to the slices of a generic point, while the two endpoints are described by the north and south pole slices. This can be formalised via the ``orbit type decomposition".
\end{example}

\subsection{The orbit type decomposition}

The orbit type decomposition allows one the characterise the quotient $M/G$ in term of conjugacy classes, $(H)$, of subgroup of $G$. First consider the decompositions 
\begin{equation}
\begin{split}
    &M = \cup_{H\subset G} M_{(H)} = \cup_i M_{i}\,,\\
    &X := M/G= \cup_{H\subset G} M_{(H)}/G = \cup_i M_{i}/G := \cup_i X_{i}\,,
\end{split}
\end{equation}
such that each $X_i$ is a connected component of $X_{(H_i)}$ and the $M_i$ are their pre-image. We can define a partial ordering of the indices $i$'s, where $i\leq j$ if $(H_i) < (H_j)$, i.e. if there exists two representatives such that $H_i \subset H_j$. Each connected component $M_i$, or $X_i$, is called  a ``\emph{stratum}" of $M$, or $X$. With these notations we have
\begin{theorem}[Orbit type stratification]
    The decompositions $M = \cup_i M_i$ and $X = \cup_i X_i$ satisfy the following conditions:
    \begin{itemize}
        \item It is \emph{locally finite}, i.e. any compact subset of $M$ (or $X$) only intersect finitely many $M_i$ (or $X_i$);
        \item \emph{It satisfies the frontier condition}, i.e. if $X_i \cap \bar{X_j}$ is not empty $\,\Rightarrow X_i \subset \bar{X_j}$ and $X_i \subset \bar{X_j}$ implies that $(H_i) \leq (H_j)$;
        \item Each $M_i$ is a smooth embedded submanifold of $M$;
        \item Each $X_i = M_i/G$ inherit a unique manifold structure for which the quotient map is a submersion;
        \item This decomposition is a \emph{stratifications}, i.e. for each $m \in X_i \subset X$, there exists an open neighbourhood $U \subset X_i$ of $m$ and a stratified space $L$ as well as a homeomorphism ${U \times \text{cone}(L) \rightarrow V \subset X}$ preserving the decomposition and restricting to diffeomorphisms between strata.
    \end{itemize}
\end{theorem}
\noindent The last condition is the most technical one but ensures that we can patch nicely the different strata $X_i$ of $X$ such that, when building $M$ from the data of the $X_i$, $M$ is a manifold.

\begin{example} We work out one non-example and one example:
    \begin{itemize}
        \item A pinched torus is not a manifold. Equipping it with a $\textrm{U}(1)$-action its quotient space is of the form $S^1$ with a distinguished stratum $X_{(U(1))}$ corresponding to the pinch. This ``stratification" satisfies all the axiom but the cone condition due to the pinched point.
        \item Let us consider the two-sphere $S^2$ with the standard $U(1)$-action. We have $X= [0,\,1]$ and the stratification structure is given by $X_{(e)} = (0,\,1)$ and two singular strata $X_{(U(1))} =\{0,\,1\}$ at each end of the interval. The singular strata correspond to the poles of the sphere.
    \end{itemize}
\end{example}
\noindent This construction allows us to state the following result:
\begin{theorem}[Principal orbit type theorem]
    Let $G\times M \rightarrow M$ be a proper group action with a connected orbit space $M/G$. There exists a unique conjugacy class $(H_{\text{prin}})$ such that $(H_{\text{prin}}) \leq (H)$ for any stabiliser group $H$ of any point $m\in M$. The corresponding orbit type  stratum $M_\text{prin}:=M_{(H_{\text{prin}})}$ is open and dense in $M$ and its quotient $X_{\text{prin}} = M_{\text{prin}}/G$ is connected, and open and dense in $X$.
\end{theorem}
\noindent The principal stratum is the only open stratum of $X$. Finally, we state that we can consider each $M_{(H)}$ as fibre bundles over $X_{(H)}$ with fibre $G/H$. More precisely we have
\begin{theorem}
    Let $H\subset G$ and $K = N_G(H)/H$, where $N_G(H)$ is the normaliser of $H$ in $G$, then there is a natural $K$ principal bundle $P_{(H)} \rightarrow X_{(H)}$ such that
    \begin{equation}
        M_{(H)} = P_{(H)} \times_K (G/H)\,.
    \end{equation}
\end{theorem}

\section{Equivariant forms}
\label{app:Equiv}

We recall some basic results concerning the construction of equivariant forms and exact equivariant forms. Let $M$ admit a proper $G$-action and $\rho: G \rightarrow \GL(V)$ be a representation of $G$ for some vector space $V$. Let $\omega$ be a $p$-form on $M$ with value in $V$, i.e. $\omega\in\Omega^p(M,\,V)$. The form $\omega$ is \emph{equivariant} if $L^*_g\omega = \rho(g) \omega$, we write $\omega \in \Omega^p(M,\,V)^G$. Infinitesimally, this implies that $\forall \,k\in\mathfrak{g} \subset \mathfrak{X}(M)$:
\begin{equation}
    \mathcal{L}_{k} \omega = \rho(k) \cdot \omega \,,
\end{equation}
to be compared with equation \eqref{eq:RedTorsionConstraints}.

\begin{theorem}Let $M$ be a reductive homogeneous space $G/H$, equivariant forms $\omega \in \Omega^p(M,\,V)^G$ are in one-to-one correspondence with $H$-invariant elements of $\Lambda^pT_{[e]}^*(G/H) \otimes V$. \label{thm:equivClass}
\end{theorem}

We show this first for equivariant functions. Let $\phi:M\rightarrow V$ be an equivariant function, i.e. $\phi \in \Omega^0(M,\,V)^G$. By definition:
\begin{equation}
    L_g (\phi(p)) = \phi(g\cdot p) = \rho(g)\cdot \phi(p)\,.
\end{equation}
As such, an equivariant function is entirely determined by its value at a single point $p$. Taking the point $p = [e]$, i.e. represented by the identity in $G$, we see immediately that $\forall h \in H$
\begin{equation}
    L_h \phi(e) = \phi(h) = \phi(e) = \rho(h) \cdot  \phi(e)\,.
\end{equation}
As such equivariant functions must be described by $H$-invariant elements of $V$. This show theorem \ref{thm:equivClass} for $p=0$.

Now, let $\omega\in \Omega^p(M,\,V)^G$ be a equivariant $p$-form with value in $V$ then
\begin{equation}
\omega(g) := \rho(g) (L^*_{g^{-1}}(\omega))(g) = \rho(g) \omega(e)(d(L_{g^{-1}})\, \cdot,\,\cdots,\,d(L_{g^{-1}})\,\cdot)
\end{equation}
which is completely specified by $\omega(e)$. One can check that it is indeed $G$-equivariant since:
\begin{equation}
\begin{split}
    (L^*_g\omega)(k) &= \omega(dL_g 
	\,\cdot\,,\dots) (gk)= \rho(gk) \omega(e)(dL_{k^{-1} g^{-1}} dL_{g}\, \cdot\,,\,\dots) \\& = \rho(g) \rho(k) \omega(dL_k\,\cdot\,,\,\dots) = \rho(g) \omega(k)\,.
\end{split}
\end{equation}
Once again, we must verify that $\omega$ is well defined, in particular we must impose that $\omega(h) = \omega(e)$. This condition implies that
\begin{equation}
    \rho(h^{-1})\omega(h) = \rho(h^{-1})\omega(e) =  \omega(e)(dh^{-1}\cdot,\,\cdots,\,dh^{-1}\cdot)
\end{equation}
Since $G/H$ is reductive, by definition, $T_eM$ admits a linear $H$-action, this is equivalent to requiring that $\omega(e) \in \Lambda^p T^*_eM \otimes V$ is an $H$-singlet, proving the correspondence between equivariant forms and $H$-invariant tensors.

A useful result for compact group $G$ states that exact $G$-equivariant forms can be expressed as the exterior derivative of another equivariant $(p-1)$-form by an averaging argument similar to \eqref{eq:AverArgInv}. Indeed, if $\omega$ is exact then $\exists\, \tilde{v}$ s.t. $d\tilde{v} = \omega$. We define
\begin{equation}
    v = |G|^{-1} \int_{G} dg \,\rho(g)^{-1} L_g^*( \tilde{v}) \,.
\end{equation} 
With these definition, $\omega = dv$ and $v$ is indeed an equivariant form. 

\section{Constraints on \texorpdfstring{$E^\flat$}{E-flat}}

\label{app:EflatConstraints}

As explained in the main text, the flat frame $E^\flat$ is somewhat constrained by the choice of principal stabiliser and the torsion constraint. In particular, we claimed that, whenever the torsion constraint reduces to $L_{K_A} K_B = \mathcal{L}_{k_A} K_B = -X_{AB}{}^C K_C$, i.e. when the gauge group is compact and the topological conditions are satisfied, $E^\flat$ must be $H$-invariant (for a specific $H$-action that we will define shortly) and it is constant on $G_g$-orbits. We will derive these results in this appendix. In order to simplify computations for this section, and without loss of generality, we will assume that we have chosen a \emph{compatible} solution to the section constraint $\mathcal{E}$ (i.e. such that $\mathbb{P}\mathcal{E} = \Theta$) and thus that
\begin{equation}
    E^\flat \in G_S \cdot \text{Stab}(\mathcal{E}_{\text{comp}}) \cdot \mathcal{S}_\mathcal{E} \,.
\end{equation}

 We emphasize that our construction in the main text yields two distinct embeddings of the abstract group $H$ in $\Ed$. The first embedding is $H_D \subset G_g \subset G_D \subset \Ed$. It is built such that it always commute with the structure group $G_S$. The second embedding is $H_\mathcal{E} \subset\GL(d')_\mathcal{E}\subset \Ed$. This one is specified by the $H$-action on the vielbein under a change of gauge ($L \rightarrow L\cdot h$), where $\mathring{e}^{-1} \rightarrow \mathring{h}^{-1}\cdot \mathring{e}^{-1}$. Importantly, these two groups need not to be conjugated to each other in $\Ed$. All the statements we make in these sections are very similar to those used in Appendix \ref{app:Equiv} concerning equivariant forms. The extra subtlety deals with here is that the frames are not equivariant but equivariant \emph{up to} a $G_S$-action and gauge transformations.

\paragraph{$H_\mathcal{E}$-invariance}
Any orbit-$G_g \cdot p$ contains a point $p$ which is invariant under the action of $H\subset G_g$. Since $H$ is compact, because the $G_g$-action is proper, the topological condition implies that the frame is equivariant under its action. We recall that under the left action by $G_g$-diffeomorphisms, the coset representative transforms as $L(g\cdot p) = g\cdot L(p) \cdot h(g,\,p)$, where $h(p,\,g)$ is the compensator (cfr. eq. \eqref{eq:cosetRepTransformations}). Applying this formula with $g = \tilde{h} \in H$ we obtain that $\forall\, \tilde{h} \in H_D$
\begin{equation}
\begin{array}{rll}
    (L_{\tilde{h}})^* (E)(p) &= {\tilde{h}}\cdot L(p)\cdot h_D\cdot E^\flat(\tilde{h}\cdot p)\cdot h_\mathcal{E}^{-1}\cdot (\mathring{e}^{-1})(p)\hspace{5mm} &\text{def. of the left action}\\
		&=h_S^{-1}\cdot  {\tilde{h}}\cdot L(p) \cdot E^\flat(p) \cdot(\mathring{e}^{-1})(p) \hspace{1cm}&\text{def. of equivariance,}
\end{array}
\end{equation}
where $h_D\in H_D \subset  G_g$ is the compensator of $\tilde{h}$ and $h_\mathcal{E}$ is the same compensator but embedded in $\GL(d')_\mathcal{E}$ rather than in $G_D$. The element $h_S \in G_S$ is there to take into account possible change of representative ($E \in G_S \backslash \Ed$). In all generality, one should worry about possible gauge transformations of the fluxes under the $H$-action. However, the topological conditions we have used ensure that we can choose a gauge for which the $p$-form potential are $H$-invariant. Since $\tilde{h}\cdot p = p$, we obtain the constraint that
\begin{equation}
     (E^\flat)^{-1}(p)\cdot (h_S \cdot h_D)\cdot E^\flat(p) = h_\mathcal{E}\,.
\end{equation}
Since, by definition, $h_S$ and $h_D$ commute this gives us, for any point of the basis, a group morphism
\begin{equation}
    \phi_p : H_\mathcal{E} \rightarrow H_D \times G_S: h_\mathcal{E}\mapsto h_D\cdot h_S = h_S \cdot h_D \,\,.
\end{equation}
This map is actually invertible on its image since $H \cong H_D$. Moreover, it is simply the $E^\flat(p)$ conjugation of $H_\mathcal{E}$. Since both $H_\mathcal{E}$ and $H_D$ are fixed, this defines a morphism, independent of $p$:
\begin{equation}
    \pi: H \rightarrow G_S : h \mapsto h_\mathcal{E} \cdot h^{-1}_D\,. 
\end{equation}
This map needs not to be invertible, we denote its image $H_S$, which is isomorphic to a subgroup of $H$.

What we have actually shown is that there is a choice of solution to the section constraint $\mathcal{E}_0 = E^\flat \mathcal{E}$ such that all $ h_{\mathcal{E}_0} \in H_{\mathcal{E}_0}$ can be decomposed as $h_{\mathcal{E}_0} = h_D \cdot h_S$ where $h_D \in H_D$ and $h_S \in H_S$ and the equality holds as matrices in $\Ed$ (and not ``up to conjugation"). Finally with the choice $\mathcal{E}_0$ of section constraint, we do have that 
\begin{equation}
    E^\flat(p) = h_D \cdot h_S \cdot E^\flat(p) \cdot h_{\mathcal{E}_0} = h_{\mathcal{E}_0}\cdot E^\flat(p) \cdot h_{\mathcal{E}_0}^{-1}
\end{equation}
at any point on the basis and for any $h \in H$. In other words, $E^\flat$ must be $H_{\mathcal{E}_0}$-invariant. Incidently, $H_{\mathcal{E}}$-invariance of the flat frame is the necessary and sufficient condition for $E$ to be globally well-defined.

\paragraph{$G_g$-equivariance}
We show that there is a choice of $G_S$-gauge for which $E^\flat$ is constant on the fibres (i.e. along $G_g$-orbits). We select again the section constraint $\mathcal{E}_0$ and get\footnote{Once again, we can safely ignore possible gauge redefinitions as the fluxes can be taken to be $G_g$-invariant for $G_g$ compact.}

\begin{equation}
\begin{array}{rll}
    (L_g)^* (E)(p) &= g\cdot L(p)\cdot h_D\cdot E^\flat(g\cdot p)\cdot h_\mathcal{E}\cdot (\mathring{e}^{-1})(p)&\text{def. of the left action}\\
		&=g_S\cdot g\cdot L(p)\cdot  E^\flat(p)\cdot (\mathring{e}^{-1})(p)\,&\text{def. of equivariance}.
    
\end{array}
\end{equation}
Once again, $g_S\in G_S$ is there to take into account possible change in the choice of representative of $E\in G_S \backslash \Ed$. Factoring out $L$, $g$ and $\mathring{e}$, we read that $\forall g_D \in G_g$, $\exists$ $g_S \in G_S$ such that
\begin{equation}
    h_D\cdot E^\flat(g\cdot p) \cdot h_\mathcal{E}^{-1} = g_S(p,\,g) E^\flat(p)\,.
\end{equation}
Using the previous result, this reduces to
\begin{equation}
    E^\flat(g\cdot p) = g_S'(g,\,p) \cdot E^\flat(p)\,.
\end{equation}
We can thus perform the gauge transformation
\begin{equation}
    (E^\flat)'(g\cdot p ) = (g_S')^{-1}(p,\,g) \cdot  (E^\flat)(g\cdot p )\,.
\end{equation}
With this choice of gauge, the flattened frame $(E^\flat)'$ is constant on the fibres and is simply a function of the basis\footnote{More mathematically inclined physicists would argue that since the frame is defined only up to $G_S$ transformation, the flat frame was already constant on each fibres. We only state that there exists a constant representative for each fibre.}.

\paragraph{Reduction to $H_S$ structure group}
 We also remark that the frame satisfying our constraint is actually a well defined section of $H_S \backslash\Ed$. One could then wonder if it is possible to further reduce the structure group to provide a consistent truncation to an $H_S$-invariant theory. This can and does happen in various cases. However, in the most generic situation, the $H_S$ intrinsic torsion associated with this new reduction of the structure group needs not being an $H_S$-singlet (since we would only quotient the torsion by $R_1 \times \mathfrak{h}_S$ and not the full  $R_1 \times \mathfrak{g}_S$). Moreover, the $H_S$ intrinsic torsion need not being constant. As such, we cannot assume that there always exists a larger truncation to an $H_S$-structure. However, this observation allows one to quickly check the existence of possible extensions to the truncated theory corresponding to $H$-structures with $G_S \supset H \supset H_S$.

\section{Kaluza-Klein calculus}
\label{app:KKCalculus}
In this appendix, we spell out how one can efficiently perform computations and checks of equations of motion in the context of KK-compactifications. In particular, we will see how the embedding tensor formalism naturally emerges from the KK ansatz. We will focus on the case $D=4$ which is the one relevant for the computations in the main text.

\subsection{Embedding tensor formalism and the tensor hierarchy}
We start by reviewing the embedding tensor formalism, which will be useful to discuss KK-compactifications of gauged supergravities. For reference, see \cite{deWit:2002vt,deWit:2005hv,deWit:2005ub}. In order to gauge a supergravity, we must use the vectors $A^M$, transforming as a representation $R_v$ of the duality group $G_D$. To gauge a subgroup $G_g\subset G_D$, one needs to specify an embedding tensor $\Theta: R_v \rightarrow \mathfrak{g}_D$:
\begin{equation}
    \Theta_M =\Theta_M{}^\alpha t_\alpha\,\hspace{1cm}\text{ s.t. }\hspace{1cm}\text{Im}(\Theta_M) = \mathfrak{g}_g\,.
\end{equation}
This allows one to define the covariant derivative
\begin{equation}
    D = d + A^M\Theta_M\,.
\end{equation}
To preserve locality of the e.o.m., $R_v$ must be a symplectic representation, i.e. it preserves a non-degenerate antisymmetric matrix $\mathbb{C}^{MN}$:
\begin{equation}
    (t_\alpha)_{MN} = (t_\alpha)_{NM}\,,
\end{equation}
where the $R_v$ indices $M$ and $N$ are lowered and raised using $\mathbb{C}_{MN}$. This tensor allows one to impose the \emph{quadratic constraints}:
\begin{equation}
\begin{split}
    &\Theta_M \,\Theta_N\, \mathbb{C}^{MN} = 0\,,\\
    &[X_M,\,X_N] = -X_{MN}{}^P X_P\,,
    \end{split}
\end{equation}
where we have defined
\begin{equation}
    X_{MN}{}^P = \Theta_M{}^\alpha (t_\alpha)_N{}^P\,.
\end{equation}
These equations imply that the embedding tensor is invariant under the action of the gauge algebra, and we get that
\begin{equation}
    \Theta_M{}^\alpha \Theta_N{}^\beta f_{\alpha\beta}{}^\gamma = - X_{MN}{}^P \Theta_P{}^\gamma\,.  
\end{equation}
We must also impose a linear constraint on the embedding tensor:
\begin{equation}
    X_{(MNP)} = 0\,,
\end{equation}
which allows one to define a unique tensor hierarchy of $p$-form, necessary to take care of magnetic charges that can appear in the gauging procedure. Finally, we define the ``intertwiner"
\begin{equation}
    Z^{M\,\alpha} = \mathbb{C}^{MN}\Theta_N{}^\alpha\hspace{1cm}\text{or equivalently}\hspace{1cm}Z^{M}{}_{PQ} = -X_{(PQ)}{}^M\,,
\end{equation}
where
\begin{equation}
    Z^M{}_{(PQ)} = Z^{M\alpha}t_{\alpha\,PQ}\,.
\end{equation}
With these definitions in hand, we can build a field strength $\mathcal{F} = F^M \Theta_M$, where
\begin{equation}
    F^M = dA^M + \frac{1}{2} A^N \wedge A^P X_{NP}{}^M\,.
\end{equation}
It turns out that, under a gauge transformation
\begin{equation}
    \delta_\xi A^M = D \xi^M = d \xi^M + X_{NP}{}^MA^N \xi^P\,,
\end{equation}
the field strength is \emph{not} covariant. This is due to the fact that $X_{MN}{}^P$ is not antisymmetric in its first two indices, which can be interpreted as the presence of magnetic gauging. To cure this, one needs to improve this field strength by adding a two-form with value in $\mathfrak{g}_D$ in the theory, $B_\alpha$, and we define the imporved field strenghts
\begin{equation}
    \mathcal{H}^M  = F^M + Z^{M\alpha} B_\alpha = F^M - X_{(PQ)}{}^M B^{PQ}\,,
    \label{eq:defHM}
\end{equation}
where $B_\alpha  = (t_\alpha)_{PQ} B^{PQ}$, in particular $B^{(PQ)}$ transform in the adjoint representation of $G$. This new field strength is gauge invariant if
\begin{equation}
    \delta_\xi B^{MN} = \mathbb{P}_{\text{adj}\,PQ}{}^{MN}(2 \xi^P \mathcal{H}^A - A^P\wedge \delta_\xi A^Q)\,.
\end{equation}
Notice that the quadratic constraint implies that
\begin{equation}
    F^M \Theta_M = \mathcal{H}^M \Theta_M\,.
\end{equation}

The addition of the 2-form requires to introduce a 1-form gauge transformation associated to it. The 1-form parameter also transforms in the adjoint representation of $G_D$ and can be written as  $\Xi_\alpha  =(t_\alpha)_{(MN)} \Xi^{MN}$. This leads to the gauge transformations
\begin{equation}
    \begin{split}
    &\delta_\Xi A^M = - Z^{M\alpha} (t_\alpha)_{(PQ)} \Xi^{PQ} = X_{(PQ)}{}^M \Xi^{PQ}\,,\\
    &\delta_{\Xi} B^{MN} = \mathcal{D}\Xi^{MN} - (\mathbb{P}_{\text{adj}})_{ST}{}^{MN}\,A^PX_{P\,QR}{}^{ST} \Xi^{QR}\,.
    \end{split}
\end{equation}
The gauge invariant 3-form field strength associated to $B^{MN}$ is
\begin{equation}
    \mathcal{H}^{MN} = DB^{MN} - (\mathbb{P}_{\text{adj}})_{PQ}{}^{MN} A^P \left(dA^Q + \frac{1}{3} X_{RS}{}^Q A^R A^S\right)\,.
\end{equation}
For $D>4$ this field strength would need to be further corrected with a 3-form, itself carrying gauge degrees of freedom and from which one can build a 4-form field strength. The procedure then repeats iteratively in what is called the \emph{tensor hierarchy} \cite{deWit:2005hv,deWit:2008gc}. 

More systematically, $p$-forms $B^{[p]}$ transform in a subrepresentation of $R_p\subset R_v^{\otimes p}$ and the embedding tensor can be used to define an \emph{intertwiner} $Y^{[p]}:R_{p+1} \rightarrow R_p$ with the properties that $Y^{[p]} \cdot Y^{[p+1]} = 0$ whenever the embedding tensor satisfies the quadratic constraints. For example, $Y^{[1]} = Z^{M\alpha} : \text{adj.} \rightarrow R_1$ and $Z^{M\alpha} \Theta_{M}{}^\beta = 0$.  Field strengths are always of the form
\begin{equation}
    \mathcal{H}^{[p]} = D B^{[p]} + \cdots + Y^{[p]} B^{[p+1]}\,.
\end{equation}
This means that $Y^{[p-1]}\,\mathcal{H}^{[p]} $ is always independent of the ($p+1$)-forms. In the context of consistent truncations, the contributions from $\mathcal{H}^{[D-2]}$ to the gauge invariant field strengths can only appears contracted with $Y^{[D-3]}$, otherwise the tensor hierarchy would not end and gauge invariance would be spoiled. We refer to \cite{Greitz:2013pua} for more details on the systematics of the tensor hierarchy.

\subsection{Kaluza-Klein twist}
To perform a KK reduction, we assume that the total manifold is $M_{\text{tot}} = M_{\text{ext}} \ltimes \Mint$ (where $\ltimes$ denotes a possible warping, as well as a fibration when KK vector fields are turned on). We also assume that the internal manifold $M_{\text{int}}$ does admit a $G_g$-action. When studying Kaluza-Klein consistent truncations, we showed in the main text that the KK vectors can be written as
\begin{equation}
    A_{KK} = A^M \Theta_M{}^\alpha \iota_{k_\alpha}\,.
\end{equation}
The $k_\alpha$ are the Killing vector associated to $\mathfrak{g}_g$ and $\iota$ denotes the interior product. The vector fields $A^M$ will be the lower-dimensional gauge fields. We will denote $k_M = \Theta_M{}^\alpha k_\alpha$ and $\iota_M = \iota_{k_M}$.

To perform the KK procedure, it proves convenient to introduce the twisting operator
\begin{equation}
    e^{A_{KK}} = \sum_n \frac{1}{n!} (A_{KK})^n\,
\end{equation}
and we define the ``KK-flat" $p$-forms $\bar{C}_{(p)}$ associated to a $p$-form $C_{(p)}$ as
\begin{equation}
    C_{(p)} = e^{A_{KK}} \bar{C}_{(p)}\,.
\end{equation}
The $p$-form $\bar{C}_{(p)}$ is the one that can be read form an expansion in terms of scalar, and lower-dimensional forms. However, in order to check equations of motion and gauge invariance, one must compute quantities of the type $dC_{(p)}$, $\star C_{(p)}$, and $B_{(p)} \wedge C_{(q)}$. We compute here how these operators commute with the twisting operator. 

We show first that \cite{Kalkman:1993zp}
\begin{equation}
    dC_{(p)} = e^{A_{KK}}\,\mathcal{D} \bar{C}_{(p)}\,,
\end{equation}
where
\begin{equation}
    \mathcal{D} = e^{-A_{KK}} d e^{A_{KK}} = d + \mathcal{H}^M \Theta_M{}^\alpha \wedge \iota_{k_\alpha} - A^M \Theta_M{}^\alpha\wedge \mathcal{L}_{k_\alpha}\,
\end{equation}
and $\mathcal{H}^M$ is defined as in \eqref{eq:defHM}.
This result is a consequence of BCH formula with $\mathcal{O}= -A_{KK}$
\begin{equation}
    e^{\mathcal{O}} d e^{-\mathcal{O}} = \sum_n \frac{1}{n!}[\mathcal{O},\,d]_n= d + [\mathcal{O},\,d] + \tfrac{1}{2} [\mathcal{O},\,[\mathcal{O}, d]] + \cdots 
\end{equation}
We compute the first terms:
\begin{equation}
\begin{array}{rl}
    [\mathcal{O},\,d] &= -A^M \iota_{k_M} d + d(A^M \iota_{k_M} )\\
    &= -A^M \iota_{k_M}d + d A^M \iota_{k_M} - A^M d\iota_{k_M}\\
    &=(dA^M \iota_{k_M} - A^M \mathcal{L}_{k_M})\,, \\ 
  {}[\mathcal{O},\,d]_2 &= -[A^M \iota_{k_M}, \,(dA^N  \iota_{k_N} - A^N \mathcal{L}_{k_N})] \\
  &= dA^M A^N \{\iota_{k_M},\,\iota_{k_N}\} - A^M A^N [\iota_{k_M},\,\mathcal{L}_{k_N}]\\
  &= 0 - A^M A^N \iota_{[k_M,\,k_N]}\\
  &= A^M\,A^N\,X_{MN}{}^P \iota_{k_P}\,,\\
  {}[\mathcal{O},\,d]_3&= [A^Q\iota_{k_Q},\, -A^M\,A^N\,X_{MN}{}^P \iota_{k_P}] \\
  &=A^Q\,A^M\,A^N\,X_{MN}{}^P \{ \iota_{k_Q},\,\iota_{k_P}\} = 0\,,
\end{array}
\end{equation}
where we have used $\mathcal{H}^M \Theta_M = F^M \Theta_M$ and that
\begin{equation}
    \mathcal{L}_{k_M} k_{N} = [k_M,\,k_N] = - X_{MN}{}^P k_P\,.
\end{equation}
You can check that $\mathcal{D}^2 = 0$, as it should. 

With the same techniques, we can compute how the KK-twisting operator commutes with the rest of the operators of Cartan calculus. The interior product is left unchanged
\begin{equation}
    e^{-A_{KK}} \iota_v e^{A_{KK}} = \iota_v\,,
\end{equation}
and the same is true for the wedge product
\begin{equation}
    e^{-A_{KK}} \left[(e^{A_{KK}}\alpha) \wedge( e^{A_{KK}}\beta)\right] = ( \alpha \wedge \beta)\,,
\end{equation}
for any two forms $\alpha$ and $\beta$. However, the Lie derivative gets corrections in $A^M$:
\begin{equation}
\begin{split}
    e^{-A_{KK}} \mathcal{L}_{v} e^{A_{KK}} &= e^{-A_{KK}}(d\iota_v + \iota_v d) e^{A_{KK}}\\
    &= [\mathcal{D}, \iota_v] = \mathcal{L}_{v} + F^M \{\iota_M,\,\iota_v\} - A^M [\mathcal{L}_M,\,\iota_v]\\
   \bar{\mathcal{L}}_v :&= \mathcal{L}_v - A^M \iota_{[k_M,\,v]}
\end{split}
\end{equation}
As such the KK twist induces an isomorphism between the superalgebras generated by $(d,\,\iota,\,\mathcal{L})$ and $(\mathcal{D},\,\iota,\,\bar{\mathcal{L}})$, which satisfy both the same supercommutation relations.

Finally, given a vielbein $e$, and hence a metric, on the total space, we introduce the operator 
\begin{equation}
    \bar{\star}=e^{-A_{KK}} \star e^{A_{KK}}\,,
\end{equation}
This operator is simply the Hodge dual with respect to the untwisted vielbein $\bar{e} = e^{-A_{KK}} e$. This is due to the fact that $\text{det}(\bar{e}) = \text{det}(e)$.

\subsection{Field strength decompositions}
We can now use the results in the main text to write down $\bar{C}_{(p)}$ in term of lower-dimensional forms. This will allow us to spell out the various non-linear field redefinitions needed to obtain gauge invariance. We stress that our computations do not depend on the dimension of the external space and are valid for any compact gauging whose uplift satsify our topological conditions. We start by showing an example to illustrate the more general construction.

\paragraph{2-forms} From the main text, we know that the piece coupling to the 1-forms $A^M$ is exact. The ansatz is thus of the form 
\begin{equation}
    \bar{C}_{(2)} = c + A^M dc_M + B^{MN} v_{MN}+ \tilde{B}^{MN} v'_{MN}\,.
\end{equation}
The last piece schematically encode further non-linear corrections while $v_{MN}$ is as yet undefined (it could be read from the ExFT 2-forms but we will show that we do not need this for the purpose of consistent truncation). Immediately, upon computing $\cD \bar{C}_{(2)}$, gauge invariance forces  $dv_{MN} = -X_{MN}{}^P dc_P$ and we can write
\begin{equation}
    \bar{C}_{(2)} = c + A^M dc_M + B^{MN} Z_{MN}{}^P c_P + \tilde{B}^{MN} c_{MN}\,,
\end{equation}
where the primitive $c_P$ of $dc_P$ is fixed by requiring equivariance. Moreover, $B^{MN}$ can only appear contracted with the intertwiner $Z_{MN}{}^P$. This intertwiner removes the need for a 3-form in order to build $\mathcal{H}^{MN}$ terms. Finally, one just needs to  fix the $\tilde{B}^{MN} c_{MN}$ term by imposing gauge invariance and we get the unique possibility
\begin{equation}
    \cD \bar{C}_{(2)} = dc + \mathcal{H}^M (dc_M + \iota_M c) + \mathcal{H}^{MN} Z_{MN}{}^P c_P 
\end{equation}
from the primitive
\begin{equation}
   \bar{C}_{(2)} = dc + A^M (dc_M ) -(B^{MN} + \frac{1}{2} A^M A^N)\,X_{MN}{}^P c_P\,. 
   \label{eq:twoformsNonLinearRedef}
\end{equation}
This computation relies on two useful identities. First, since $c_P$ is an equivariant scalar, $\iota_M dc_P = \mathcal{L}_M c_P = -X_{MP}{}^N c_N$, and second 
\begin{equation}
    \left[X_{MP}{}^R X_{[NR]}{}^Q\right] = \frac{1}{3} \left[X_{MN}{}^R X_{(RP)}{}^Q\right]_{[MNP]}\,.
\end{equation}
The result \eqref{eq:twoformsNonLinearRedef} is universal for all the consistent truncation we consider in this paper (i.e. the ones such that the generalised sections are made of (co-)closed equivariant poly-forms) and does not depends on the number of external dimensions.

\paragraph{Higher $p$-forms}
The non-linear corrections needed to make higher $p$-forms gauge-invariant can be obtained in the same spirit, iterating on the rank of external forms contributions. This is the strategy we will use to study the corrections to $C_{(4)}$ in type IIB. Notice that the highest-rank $(D-2)$-forms must always appear contracted with $Y^{[D-3]}$ in order remove the need for any unphysical $(D-1)$-form in the theory.

\subsection{The \texorpdfstring{$\Es$}{E-7(7)}-ExFT/type IIB dictionary}

Any $p$-form splits into various contributions, organised by their rank in the external space. For example
\begin{equation}
    \bar{C}_{(p)} = c + A^M c_M + B^{MN} c_{MN} + \cdots
\end{equation}
where $c$ is the scalar contribution, read from the generalised metric $\mathcal{M}_{MN}(x,\,Y^M)$. For the precise expressions in the type IIB context, see \cite{Inverso:2016eet}. The vector contributions $A^M c_M$ are given, before any non-linear field redefinitions, by reading the appropriate components of the sections $A^M K_{M\,(p-1)}$. In order to preserve gauge invariance, we must then add contributions from the 2-forms as well as a series of corrections non-linear in the lower-dimensional supergravity fields, represented here by ``$\cdots$".

\paragraph{The 2-forms}
The non-linear field redefinitions for the higher-dimensional 2-forms do not depend on $D$. The resulting type IIB 2-forms are
\begin{equation}
\label{eq:2formDecomposition}
    \bar{\mathbb{B}}^\alpha = b^\alpha + A^M db_{M}^\alpha - \left(B^{MN}+ \frac{1}{2} A^M A^N\right) X_{MN}{}^{P} b_P^\alpha\,.
\end{equation}
There are 3 types of contributions coming from the sections and the generalised metric. The first one $b^\alpha$ is the scalar contribution obtained from the generalised metric. The second one $b_M^\alpha$ is read from the 1-form piece of the section  $b^\alpha = K_{M\,(1)}{}^\alpha$. Finally, its primitive, $b_P$, is uniquely fixed by requiring its equivariance. 

The corresponding fluxes can be read in a very compact manner as
\begin{equation}
    \cD\bar{\mathbb{B}}^\alpha = db^\alpha + \mathcal{H}^M \left(db^\alpha_M + i_M b^\alpha \right) + \mathcal{H}^{MN} Z_{(MN)}{}^Pb^\alpha_P\,.
\end{equation}

\paragraph{The 4-form} The 4-form is more subtle. Indeed, the gauge invariant object in type IIB supergravity is not $F_{(5)} = dC_{(4)}$ but $\tilde{F}_{(5)} = dC_{(4)} + \frac{1}{2} \epsilon_{\alpha\beta}\mathbb{B}^\alpha d \mathbb{B}^\beta$. This implies that building a 4-form such that $dC_{(4)}$ is gauge invariant does not imply that $\tilde{F}_{(5)}$ is. In particular, it receives contributions of the form $\frac{1}{2} \epsilon_{\alpha\beta} A^M b_M^\alpha db^\beta$ which must be compensated. For the case at hand, the field redefinitions are
\begin{equation}
\begin{split}
    \bar{C}_{(4)} =& c + A^M \left(dc_M + \frac{1}{2} \epsilon_{\alpha\beta}  b^\alpha db_M^\beta\right)\\
    &- \left(B^{MN}+\frac{1}{2} A^M A^N\right) X_{MN}{}^P \left(c_P+ \frac{1}{2} \epsilon_{\alpha\beta}  b^\alpha b_M^\beta\right)\,,
\end{split}
\end{equation}
where $dc_M$ is read from the three form components of the sections. The contributions from its primitive, $X_{MN}{}^P c_P$, is fixed by equivariance. From this we obtain the improved 5-form:
\begin{equation}
\begin{split}
    \bar{\tilde{F}}_5 =& (1+\bar{\star})\left(dc +\frac{1}{2}\epsilon_{\alpha\beta} b^\alpha db^\beta\right)\\
    &+\mathcal{H}^M \left(dc_M + \iota_M c + \epsilon_{\alpha\beta} b^\alpha db_M^\beta + \frac{1}{2}\epsilon_{\alpha\beta} b^\alpha \iota_M b^\beta  \right)\\
    &-\mathcal{H}^{MN}X_{MN}{}^P(c_P + \epsilon_{\alpha\beta} b^\alpha b^\beta_P).
\end{split}
\end{equation}
We have included the contributions dual to the internal scalar contributions in order for $\bar{\tilde{F}}_5$ to be self-dual, hence the presence of $\bar\star$ in the first line of the previous equation.

\bibliography{reference}

\end{document}